\documentclass[a4paper,11pt]{article}

\usepackage[utf8]{inputenc} %Czech and Slovak symbols
\usepackage{lmodern}
\usepackage[T1]{fontenc}
\usepackage{textcomp}
\usepackage{natbib} %standard package for bibliography
\usepackage{amsmath,amsthm,amssymb} %better typesetting of mathematical expressions, theorems, additional symbols
\usepackage{booktabs} %journal-type tables
\usepackage{caption} %takes care of tables' and figures' captions
\usepackage{rotating} %landscape tables (floating)
\usepackage{longtable} %for long tables over one page
\usepackage{lscape} %landscape tables (floating)
\usepackage{tabularx}
\usepackage{dcolumn}
\usepackage{changepage}
\usepackage{graphicx}
\usepackage{fullpage} %narrower margins
\usepackage{multicol} %text such as bibliohraphy in more columns
\usepackage{comment} %comment blocks
\usepackage{authblk} %author blocks
\usepackage{subfigure} %subfigures
\usepackage{setspace} %doublespacing
\usepackage{hyperref} %hyperlink support
\usepackage{threeparttable}
\usepackage{float}
\usepackage{pdflscape}
\usepackage{geometry}
\hypersetup{pdfstartview={XYZ null null 1.00},bookmarksnumbered,hypertexnames=false} %zoom: default screen; numbers for bookmarks
\usepackage{color,soul}
\soulregister\cite{7}
\soulregister\citet{7}
\soulregister\citep{7}
\soulregister\autoref{7}
\soulregister\url{7}

\newcolumntype{d}[1]{D{.}{. }{#1}}

\begin{document}
%\onehalfspacing
\doublespacing

\hyphenation{elas-tic-i-ties E-con-o-mies} %hyphenation of selected words
\renewcommand{\sectionautorefname}{Section} %capital S at the beginning
\newcolumntype{L}[1]{>{\raggedright\let\newline\\\arraybackslash\hspace{0pt}}m{#1}}  % define column width in tabular as L{3cm}
\newcolumntype{C}[1]{>{\centering\let\newline\\\arraybackslash\hspace{0pt}}m{#1}}
\newcolumntype{R}[1]{>{\raggedleft\let\newline\\\arraybackslash\hspace{0pt}}m{#1}}
\def\sym#1{\ifmmode^{#1}\else\(^{#1}\)\fi} %shortcut for Stata tables
\clubpenalty 9999 %no orphans (typographic properties)
\widowpenalty 9999 %no widows (typographic properties)

\title{The Elasticity of Substitution between Native and Immigrant Labor: A Meta-Analysis\thanks{An online appendix with data and code is available at \url{https://meta-analysis.cz/migrant/}. Corresponding author: Klara Kantova, \url{klara.kantova@fsv.cuni.cz}. Kantova acknowledges support from the Charles University, project GA UK No. 152222 and No. 378225, from the European Union's Horizon 2020 research and innovation program under the Marie Sk\l{}odowska-Curie grant agreement No. 870245, and from the Charles University Research Centre program No. 24/SSH/020. Havranek, Irsova, and Schwarz acknowledge support from the Czech Science Foundation (grant 24-11583S).}}

\author[a]{Klara Kantova}
\author[a,b,c]{Tomas Havranek}
\author[d]{Zuzana Irsova}
\author[d]{Jiri Schwarz}
\affil[a]{Institute of Economic Studies, Charles University, Prague}
\affil[b]{Centre for Economic Policy Research, London}
\affil[c]{Meta-Research Innovation Center at Stanford}
\affil[d]{Anglo-American University, Prague}

\date{July 23, 2026}
\maketitle
\thispagestyle{empty}
\enlargethispage{4\baselineskip}

\vspace{-1.0 cm}

\begin{abstract}\scriptsize
This paper presents the first comprehensive meta-analysis of the elasticity of substitution between native and immigrant labor, drawing on 1,091 estimates from 41 studies. We find strong evidence of selective reporting: less precise estimates are systematically associated with lower reported elasticities. Correcting for this bias using meta-regression and selection methods raises the implied elasticity from about 13 to about 22, implying about 40\% less relative-wage pressure from immigration than uncorrected results suggest. Bayesian and frequentist model averaging show that heterogeneity is driven mainly by geographic scale, data granularity, and whether the sample is restricted to low-experience workers, while the choice between log mean wages and mean log wages plays a secondary role. Our best-practice estimates, which net out publication bias and prioritize the most granular data, imply an elasticity of about 17 in our baseline regional specification, lower than implied by a simple bias correction but substantially higher than the uncorrected mean.  
\end{abstract}

\vspace{3pt}
\begin{tabular}{p{0.2\hsize}p{0.7\hsize}} %0.15
\textbf{Keywords:} & elasticity of substitution, immigration, native labor, meta-analysis, publication bias, selective reporting \\
\end{tabular}

\smallskip

\begin{tabular}{p{0.2\hsize}p{0.4\hsize}}
\textbf{JEL Codes:} &  J15, J61, C83 \\
\end{tabular}

\newpage

\section{Introduction}
\label{sec:intro}

The elasticity of substitution between native and immigrant labor, $\sigma$, is the structural parameter that governs how an inflow of foreign workers feeds through to native wages and employment. When the two groups are close substitutes, immigrants compete with comparable natives and put downward pressure on their pay. When they are complements, an inflow can raise native productivity and earnings. Most quantitative claims about who gains and who loses from immigration rest, explicitly or implicitly, on a value of $\sigma$. Pinning it down is therefore the empirical question on which the wage and employment effects of immigration ultimately turn.

Empirical estimates of $\sigma$ span a wide range, and the field has remained divided on their interpretation since the exchange between \citet{borjas2003labor} and \citet{Card2001} over whether immigrants compete directly with comparable natives or fill complementary roles. \citet{CardMariel} found little sign that the Mariel influx depressed native wages in Miami, a reading \citet{Borjas2017} disputed using narrower subgroup analyses. \citet{ottaviano2012rethinking} and \citet{Manacorda2012} place the parameter at imperfect but economically meaningful substitutability. \citet{Dustman2016} and \citet{Edo2019} trace the disagreement to method: \citet{Edo2019} sorts the literature into structural education-experience cell models, spatial correlation designs, and national skill-cell approaches, each of which can deliver a different $\sigma$ from the same data. \citet{Kerr2011} adds that the effects are uneven across workers, falling hardest on less-educated natives and earlier immigrant cohorts. With estimates this dispersed and so closely tied to research design, a quantitative synthesis is needed to recover what the literature collectively implies.

The theoretical drivers of these elasticities are equally diverse. \citet{George2011} highlights that wage impacts depend on the interplay between substitution across skill levels and substitution within skill cells. While some US evidence suggests skilled immigrants and natives may be perfect substitutes \citep[e.g.,][]{Borjasetal2008}, \citet{Peri2011} and \citet{PERI20111} argue they are imperfect substitutes because they specialize in different tasks as immigrants often concentrate on quantitative and analytical skills, while natives focus on interactive and communication-based roles. Yet, despite these theoretical insights, reported elasticities in primary studies range from 6 to over 500.

Primary studies often cannot estimate the structural elasticity directly. Data limitations, model complexity, or identification concerns often lead researchers to estimate related labor market effects, such as the wage impacts of immigration, without recovering the elasticity itself. In such cases, the elasticity is frequently parameterized using values from existing studies, including meta-analytic syntheses of related labor-market elasticities \citep{elminejad2023labor}. However, as \citet{aubry2026does} show in their comprehensive meta-analysis of nearly 3,000 reduced-form wage estimates, methodological choices significantly shape reported results. If these parameters are miscalibrated, it can distort empirical conclusions and policy simulations. This highlights the need for a systematic synthesis of the empirical literature that corrects for publication bias and explains the underlying heterogeneity.

Meta-analysis provides a formal framework for distilling patterns from such fragmented findings \citep{stanley2001wheat}. While the work by \citet{aubry2026does} offers a vital assessment of total wage impacts, it does not isolate the structural elasticity of substitution. Our paper addresses this gap by conducting the first comprehensive meta-analysis of elasticity based on 1,091 estimates from 41 primary studies. In doing so, we follow a growing tradition of using meta-analysis to calibrate fundamental structural parameters, such as the capital-labor elasticity of substitution \citep{gechert2022measuring} and the Armington elasticity \citep{bajzik2020estimating}. We apply a wide array of graphical and meta-regression techniques to recover bias-corrected estimates that better approximate the underlying elasticity. We also systematically investigate sources of heterogeneity using both Bayesian and frequentist model averaging. Following \citet{havranek2024publication}, we focus on the negative inverse elasticity ($-1/\sigma$), which we hereafter refer to as the coefficient. Reported elasticities of roughly 6 to 500 correspond to coefficients between about $-0.17$ and $-0.002$. This parameter is the most commonly reported and is typically derived from specifications that use relative labor supply as the main source of variation. To ensure comparability within the meta-regression framework, we exclude estimates reported directly as the elasticity rather than as the inverse-elasticity coefficient ($-1/\sigma$), because such estimates typically lack the standard errors required for meta-regression or would require inversions that violate the assumptions of publication bias tests \citep{havranek2024publication}. This screen is applied estimate by estimate: a study whose results are reported only as directly stated elasticities (e.g., \citealp{angioloni2022liberalizing}) is excluded in full, whereas a study that also reports usable inverse-elasticity coefficients is retained for those estimates and so still appears in our list of included studies (\autoref{tab:app:studies}).\footnote{Because our meta-analytic outcome is $\theta = -1/\sigma$, the transformation is nonlinear: large values of $\sigma$ are compressed into a narrow interval near zero, so small rounding or reporting differences in directly reported elasticities (and their standard errors) can translate into non-negligible differences in $\theta$ and its precision.}

Three main findings emerge. First, the literature is consistent with substantial selective reporting. We find that less precise estimates are systematically associated with more negative reported values (lower substitutability). Because the correlation between the reported coefficients and the inverse square root of the sample size is negligible ($r = 0.016$), this pattern reflects selective reporting based on precision rather than a "small-study" effect driven by sample size \citep{irsova2025spurious}. After correcting for this bias, the implied elasticity rises from an uncorrected average of 13 to approximately 22. This increase is economically meaningful because a higher elasticity implies a smaller relative-wage response to a given change in relative labor supply. Under the standard inverse-elasticity relationship, a 10\% increase in immigrant labor relative to native labor lowers immigrant wages relative to native wages by about 0.77\% when $\sigma = 13$, but only by about 0.45\% when $\sigma = 22$. The correction thus reduces the implied relative-wage pressure by roughly 40\%.

Second, our results demonstrate that much of the variation in reported estimates is driven by identifiable data features. Reported substitutability tends to be higher in studies with larger immigrant shares and a finer division of the labor market into worker groups, or \textit{cells}, defined by combinations of education, experience, region, gender, and time. Specifically, as researchers increase the number of cells, reported substitutability rises. This suggests that low-resolution models with fewer cells may aggregate distinct labor markets, potentially masking the true degree of competition. Conversely, studies using annual data frequency or hourly wages report lower substitutability. Building on these findings, we calculate a best-practice estimate that nets out publication bias while prioritizing the most granular data available, with hourly wages and annual data frequency. We identify an elasticity of approximately 17 for regional models and 8 for national models.

Third, we revisit the methodological debate regarding the transformation of the dependent variable, i.e., the log of mean wages versus the mean of log wages. While the raw descriptive averages mirror the debate between \citet{borjas2012com_est} and \citet{ottaviano2012rethinking}, our analysis reveals that this choice is a significant, albeit secondary, driver of heterogeneity. This suggests that the wage definition is a systematic factor that researchers must calibrate alongside other data features, and does not invalidate previous findings.

The remainder of the paper proceeds as follows. \autoref{sec:data} describes the data collection process and variables, providing a detailed definition of our data granularity measures. \autoref{sec:pubbias} assesses and accommodates potential selective reporting using both linear and advanced techniques. \autoref{sec:het} explores the drivers of heterogeneity using Bayesian and frequentist model averaging, including an analysis of the wage-definition debate. This section also derives our best-practice estimates, demonstrating how conditioning on the most granular data available shifts the implied elasticity. \autoref{sec:concl} contextualizes the findings within the broader immigration policy debate and concludes.

\section{Data}
\label{sec:data}

The elasticity of substitution between immigrant and native labor is mathematically defined as:
\[
\sigma = -\, \frac{d \log \left( \frac{L_i}{L_n} \right)}{d \log \left( \frac{w_i}{w_n} \right)}
\]
where $L_i$ and $L_n$ are the quantities of immigrant and native labor employed, and $w_i$ and $w_n$ are their respective wages. This definition shows how the ratio of immigrant to native labor changes in response to changes in their relative wages \citep{borjas2010labor}. 

Standard economic theory implies a nonnegative elasticity of substitution between immigrant and native workers: when immigrants become relatively cheaper, their relative employment should not fall. Negative estimates of $\sigma$ are possible in empirical work, but they are difficult to interpret, since they imply that a decline in immigrants' relative wage would be associated with lower relative immigrant employment. \cite{kearney1997estimating} and \cite{bowles1970aggregation} disregard such values as estimation artifacts, though \cite{ribo2020restrictions} suggests they may indicate extreme complementarity. 

\subsection{Collecting the Elasticity Dataset}

To gather our estimates, we searched Google Scholar using combinations of the keywords "elasticity of substitution," "immigrant," "labor," and "native." By January 31, 2025, we reviewed 960 records and collected 1,091 estimates from 41 studies. In \autoref{app:Data}, we provide the detailed list of included studies (\autoref{tab:app:studies}) and the PRISMA diagram (\autoref{fig:PRISMA}), which outlines our specific inclusion and exclusion criteria.

Following \cite{havranek2024publication}, we focus on the negative inverse elasticity, $-1/\sigma$, as relative labor supply variations are generally more exogenous than wage changes. The baseline specification is the Constant Elasticity of Substitution (CES) production function:
\begin{equation}\label{eq:CES}
 Y = [\alpha(aL_{n})^{\rho} + (1-\alpha)(bL_{i})^{\rho}]^{\frac{1}{\rho}},
\end{equation}
where $L_n$ and $L_i$ are native and immigrant labor, $a$ and $b$ are factor-augmenting technologies, and $\alpha$ is the native labor share. From this, $\sigma = \frac{1}{1-\rho}$. The estimated equation in primary studies is:
\begin{equation}\label{eq:est}
\ln\left(\frac{w_{n}}{w_{i}}\right) = A - \frac{1}{\sigma} \ln \left(\frac{L_{n}}{L_{i}}\right),
\end{equation}
where $w_{n}/w_{i}$ is the native wage premium, $L_{n}/L_{i}$ is the relative labor supply, and $A$ captures other factors, such as skill-specific technology shocks and fixed effects. This follows the structural approach widely adopted in the literature \citep[e.g.,][]{borjas2003labor, ottaviano2012rethinking, Manacorda2012}. In this log-linear specification, the coefficient on relative labor supply, $-1/\sigma$, represents the inverse elasticity of substitution. A more negative coefficient implies that immigrants and natives are less substitutable, as an increase in immigrant supply leads to a larger drop in their relative wages (i.e., a higher native wage premium). Of the 41 studies in our dataset, 26 adopt the log of mean wages, while 18 use the mean of log wages.\footnote{The counts sum to more than 41 because three studies use both measures.} Among the latter, key examples include \cite{Borjas2010}, \cite{borjas2012com_est}, and \cite{Card2009}, who argue that this specification better captures individual-level wage variation. In contrast, \cite{ottaviano2012rethinking} use log of mean wages, which \cite{borjas2012com_est} critiques for overstating negative impacts.

Initially, we collected a comprehensive set of potential moderators. However, to ensure robust identification and avoid misleading inference, we excluded variables with insufficient variation (i.e., means below 0.03 or above 0.97). This process resulted in a final set of 27 moderator variables spanning data characteristics, structural variations, estimation techniques, fixed effects, and publication characteristics. One of our key variables is the number of \textit{cells} used in the structural estimation, which represents the product of education and experience groups (and sometimes region, year, and/or gender). This variable serves as a measure of data granularity. For instance, the national skill-cell approach popularized by \citet{borjas2003labor} typically uses 192 cells (8 experience groups $\times$ 4 education groups $\times$ 2 genders $\times$ 3 census years). In the data, we also include the percentage of foreign-born population in the respective country, the number of citations, journal impact factor, and publication year. \autoref{tab:tab1} presents summary statistics for the sample, showing simple means without correcting for publication bias, and provides insights into the elasticity of substitution between immigrant and native labor across various subsamples and characteristics. In the Appendix, specifically \autoref{tab:descr}, we define these variables along with their simple mean, standard deviation, and mean weighted by the inverse of the number of estimates reported per study.\footnote{Unlike \autoref{tab:tab1}, which covers all 1,091 collected estimates, the summary statistics in \autoref{tab:descr} are computed on the 1,087 observations used in the heterogeneity analysis. Four estimates from one multi-country study are excluded there because country-specific moderators, such as the immigrant share, cannot be assigned to them (see \autoref{sec:het}).}

As noted in the Introduction, the variation in these estimates is not strongly correlated with sample size. Specifically, the correlation between the reported coefficients and the inverse square root of the sample size is near zero ($r = 0.016$). This suggests that the observed reporting pattern is linked more closely to estimate precision than to a traditional "small-study" effect driven by sample size. We also examined the impact of potential outliers by applying winsorization at the 1\%, 2.5\%, and 5\% levels. Since the results remained stable across all levels, no winsorization was ultimately applied. 

The overall mean of the reported coefficients ($-1/\sigma$) in our sample is $-0.075$, which corresponds to an implied elasticity of 13.3. To account for the influence of studies reporting multiple results, we also compute a weighted mean where each observation is weighted by the inverse of the number of estimates per study. This weighted mean coefficient is slightly more negative at $-0.077$, implying an elasticity of 12.9.\footnote{The column $\sigma$ in \autoref{tab:tab1} depicts the implied elasticity computed as $\sigma = -1/mean$ using the coefficient values prior to rounding.} Consistent with the imperfect-substitution reading of \citet{ottaviano2012rethinking}, this magnitude indicates a significant degree of imperfect substitutability. Functionally, it implies that relative wages respond only modestly to shifts in relative labor supply. Under a structural parameter of 13, a 10\% increase in immigrant labor relative to native labor lowers the immigrant-to-native wage ratio by approximately 0.77\% ($1/13 \cdot 10\%$). However, substantial variation exists across different subsamples, suggesting that factors such as data characteristics, worker type, estimation techniques, and publication sources significantly influence the reported coefficients.

The choice of dependent variable transformation, specifically the log of mean wages versus the mean of log wages, represents a significant methodological divide in the literature. Theoretically, due to Jensen’s Inequality, $E[\log(X)] \leq \log(E[X])$, using the log of mean wages may overstate the wage impact of immigration by downplaying within-group variation, potentially leading to a downward bias in the estimated elasticity of substitution.\footnote{In practical terms, the difference is very small when wage dispersion within a cell is low, but it becomes economically meaningful as dispersion rises. For example, the gap is about 0.5\% at low dispersion (with the coefficient of variation $CV = 0.1$), around 3\% at moderate dispersion ($CV = 0.25$), and can exceed 10\% when dispersion is high ($CV = 0.5$).} This choice is central to the debate between \citet{borjas2012com_est} and \citet{ottaviano2012rethinking}, where the transformation is argued to be a primary driver of differing elasticity estimates. While we revisit this debate formally in our heterogeneity analysis, our data suggest that this technical choice is often bundled with other methodological preferences. Specifically, the "Borjas vs. Ottaviano" divergence likely stems from a broader modeling philosophy, where the preference for national-level structural models often coincides with specific wage definitions, rather than the wage transformation in isolation.

\begin{figure}[htbp]
\centering
\caption{Box plot of negative inverse elasticity across studies}
\includegraphics[width=0.8\linewidth]{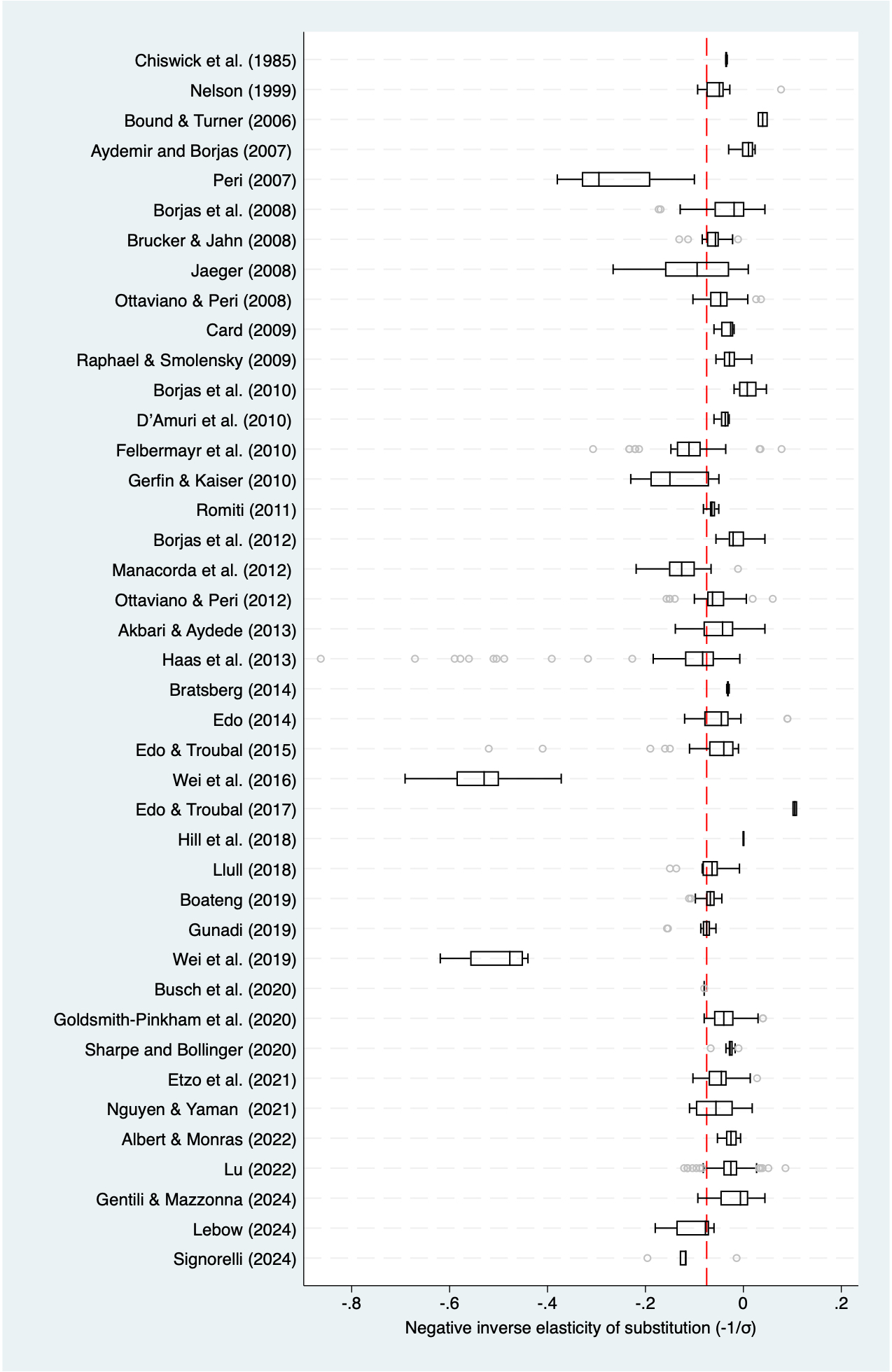}
\label{fig:study}
\begin{tabular*}{1\hsize}{@{\hskip\tabcolsep\extracolsep\fill}cccc}
\multicolumn{4}{p{1\hsize}}{\scriptsize \emph{Notes:} $-1/\sigma = \theta$, i.e., the reported coefficient. The length of each box represents the interquartile range (P25--P75), and the dividing line inside the box is the median value. The whiskers represent the highest and lowest data points within 1.5 times the range between the upper and lower quartiles. The dots show the outlying estimates with extreme values stacked at the values denoted as "outliers". The red vertical line presents the unweighted mean of all reported coefficients ($-0.075$).}
\end{tabular*}
\end{figure}

\autoref{fig:study} shows a box plot of the reported coefficients across individual studies, sorted by publication year, illustrating variations across studies (variation across countries is depicted in \autoref{fig:country} and the summary statistics in \autoref{tab:tab2_countr}). Notably, studies by \cite{wei2016imperfect} and \cite{wei2019substitution} focus on farm workers, explaining their distinct reported coefficients. The distinctiveness in these coefficients arises from the specific nature of farm work, which often involves labor-intensive tasks requiring particular physical endurance and skills typically found among younger workers \citep{wei2016imperfect, wei2019substitution}.

%%%%TABLE 1: summary stats
\begin{singlespace}
\begin{table}[H]\scriptsize
\begin{center}
\caption{Summary statistics for different subsets of the literature\label{tab:tab1}}
\begin{tabular*}{1\hsize}{@{\hskip\tabcolsep\extracolsep\fill}l*{10}{c}@{}}
\toprule
                         &             & \multicolumn{4}{c}{Unweighted}                           & \multicolumn{4}{c}{Weighted}                             \\\cline{3-6}\cline{7-10}
Variable                 & N & Mean   & $\sigma$ & \multicolumn{2}{l}{95\% conf. int.} & Mean   & $\sigma$ & \multicolumn{2}{l}{95\% conf. int.} \\
    \midrule
    
All estimates               & 1,091       & -0.075     & 13.3      & -0.08           & -0.07 & -0.077   & 12.9      & -0.08           & -0.07 \\
    \midrule
\multicolumn{3}{l}{\textit{    Data characteristics}}        &           &                  &                  &        &           &                  &                  \\
Annual frequency data       & 488         & -0.11     & 9.2       & -0.12           & -0.10 & -0.11   & 9.4       & -0.12           & -0.09 \\
Lower frequency data        & 603         & -0.05    & 20.8      & -0.05           & -0.04 & -0.05   & 22.1      & -0.05           & -0.04 \\
\midrule

\multicolumn{3}{l}{\textit{    Structural variation}}        &           &                  &                  &        &           &                  &                  \\
All workers                 & 603         & -0.09      & 11.4      & -0.10           & -0.08 & -0.09    & 10.9      & -0.10           & -0.08 \\
Full-time workers           & 488         & -0.06      & 16.8      & -0.06           & -0.05 & -0.06    & 17.3      & -0.06           & -0.05 \\
High level of experience    & 42          & -0.06      & 17.0      & -0.07           & -0.05 & -0.04    & 22.6      & -0.06           & -0.03 \\
Low level of experience     & 79          & -0.12      & 8.4       & -0.16           & -0.08 & -0.12    & 8.3       & -0.16           & -0.08 \\
High level of education     & 106         & -0.08      & 13.0      & -0.09           & -0.06 & -0.05    & 19.8      & -0.07           & -0.04 \\
Low level of education      & 171         & -0.06      & 15.5      & -0.07           & -0.06 & -0.07    & 13.7      & -0.08           & -0.06 \\
English                     & 735         & -0.07      & 13.6      & -0.08           & -0.07 & -0.09    & 11.7      & -0.10           & -0.07 \\
Non-English                 & 356         & -0.08      & 12.8      & -0.09           & -0.07 & -0.06    & 15.6      & -0.07           & -0.05 \\
Top 6 languages             & 916         & -0.07      & 14.4      & -0.08           & -0.06 & -0.08    & 12.7      & -0.09           & -0.07 \\
Male                        & 449         & -0.05      & 20.1      & -0.06           & -0.04 & -0.04    & 24.8      & -0.05           & -0.03 \\
Female                      & 98          & -0.07      & 14.5      & -0.08           & -0.05 & -0.09    & 11.6      & -0.11           & -0.06 \\
Both                        & 544         & -0.10      & 10.3      & -0.11           & -0.09 & -0.11    & 9.2       & -0.12           & -0.10 \\
Farmers                     & 34          & -0.52      & 1.9       & -0.55           & -0.49 & -0.52    & 1.9       & -0.55           & -0.49 \\
Non Farmers                 & 1,057       & -0.06      & 16.5      & -0.07           & -0.06 & -0.05    & 18.3      & -0.06           & -0.05 \\
North America               & 710         & -0.07      & 14.0      & -0.08           & -0.06 & -0.08    & 12.0      & -0.09           & -0.07 \\
Other region                & 381         & -0.08      & 12.3      & -0.09           & -0.07 & -0.07    & 14.8      & -0.08           & -0.06 \\

\midrule

\multicolumn{4}{l}{\textit{    Estimation characteristics}}              &                  &                  &        &           &                  &                  \\
DV: Log of mean wages            & 865         & -0.08      & 12.2      & -0.09           & -0.07 & -0.11    & 9.1       & -0.12           & -0.10 \\
DV: Mean of log wages          & 226         & -0.05      & 20.3      & -0.05           & -0.04 & -0.03    & 36.1      & -0.03           & -0.02 \\
Annual wage                 & 97          & -0.05      & 21.8      & -0.05           & -0.04 & -0.02    & 60.8      & -0.03           & -0.01 \\
Monthly wage                & 107         & -0.11      & 8.8       & -0.14           & -0.08 & -0.09    & 10.7      & -0.12           & -0.07 \\
Weekly wage                 & 486         & -0.05      & 19.2      & -0.06           & -0.05 & -0.06    & 15.4      & -0.07           & -0.06 \\
Daily wage                  & 49          & -0.06      & 17.8      & -0.06           & -0.05 & -0.05    & 20.3      & -0.06           & -0.04 \\
Hourly wage                 & 352         & -0.11      & 9.5       & -0.12           & -0.09 & -0.12    & 8.7       & -0.13           & -0.10 \\
National                    & 846         & -0.07      & 13.4      & -0.08           & -0.07 & -0.08    & 12.6      & -0.09           & -0.07 \\
Regional                    & 245         & -0.08      & 13.1      & -0.09           & -0.07 & -0.07    & 14.0      & -0.08           & -0.06 \\
OLS                         & 794         & -0.07      & 14.1      & -0.08           & -0.07 & -0.08    & 13.2      & -0.09           & -0.07 \\
IV                          & 297         & -0.08      & 13.0      & -0.09           & -0.06 & -0.08    & 12.1      & -0.09           & -0.07 \\

Time fixed effects          & 777         & -0.09     & 11.7      & -0.09           & -0.08 & -0.09   & 10.7      & -0.10           & -0.08 \\
Person fixed effects        & 249         & -0.14     & 7.2       & -0.16           & -0.12 & -0.12   & 8.1       & -0.15           & -0.10 \\
Skill fixed effects         & 955         & -0.07     & 14.3      & -0.08           & -0.06 & -0.07   & 14.4      & -0.08           & -0.06 \\
\midrule

\multicolumn{4}{l}{\textit{    Publication characteristics}}             &                  &                  &        &           &                  &                  \\
Published                   & 781         & -0.07      & 14.2      & -0.08           & -0.06 & -0.06    & 15.4      & -0.07           & -0.06 \\
Unpublished                 & 310         & -0.09      & 11.5      & -0.10           & -0.07 & -0.12    & 8.6       & -0.14           & -0.10 \\
Top 5 journals              & 64          & -0.04      & 26.0      & -0.05           & -0.03 & -0.04    & 26.3      & -0.05           & -0.03 \\
\bottomrule 
\end{tabular*}
\begin{tabular*}{1\hsize}{@{\hskip\tabcolsep\extracolsep\fill}cccccccccc}
\multicolumn{10}{p{1\hsize}}{\scriptsize \emph{Notes:} The exact definition of the variables is available in \autoref{tab:descr}. Weighted = coefficients are weighted by the inverse of the number of estimates reported per study. The column $\sigma$ depicts implied elasticity computed as $\sigma = -1/mean$ using the coefficient values prior to rounding. Thus, it may not perfectly correspond to the rounded mean values presented in the table. $N$ is the number of observations per sample. No winsorization needed.}
\end{tabular*}
\end{center}
\vspace{-0.4cm}
\end{table}
\end{singlespace}

\newpage
\section{Publication Selection Bias}
\label{sec:pubbias}

Publication selection bias, or reporting bias, poses a major challenge in meta-analysis, particularly when studies with significant or expected results are more likely to be reported than those with null or unexpected findings. Such selective reporting can lead to systematic exaggeration of the evidence. It can stem both from the selective publication of entire studies and from the specification and estimation choices researchers make within them, a form of \textit{p}-hacking. A growing methodological literature emphasizes distinguishing these channels from one another and from a genuine small-study effect \citep{havranek2024publication, irsova2025spurious}. With more than a thousand estimates of a single structural parameter, our setting is well suited to distinguishing selection linked to reported precision from a conventional small-study effect. To identify and accommodate potential bias, we apply a wide array of graphical and meta-regression techniques \citep{irsova2024meta} to recover bias-corrected estimates that better approximate the underlying elasticity.

A funnel plot, developed by \citet{egger1997bias}, is a scatter diagram that plots the size of a study’s effect estimate horizontally against its inverse standard error (a proxy for precision) on the vertical axis. The funnel plot is a widely used visual tool for detecting publication bias \citep{stanley2010picture}. In the absence of publication bias, the plot should be symmetric around the mean estimate. 

\begin{figure}[H]\centering
\caption{Funnel plot suggests asymmetry among coefficients}
\label{fig:funnel}
\includegraphics[width=1\linewidth]{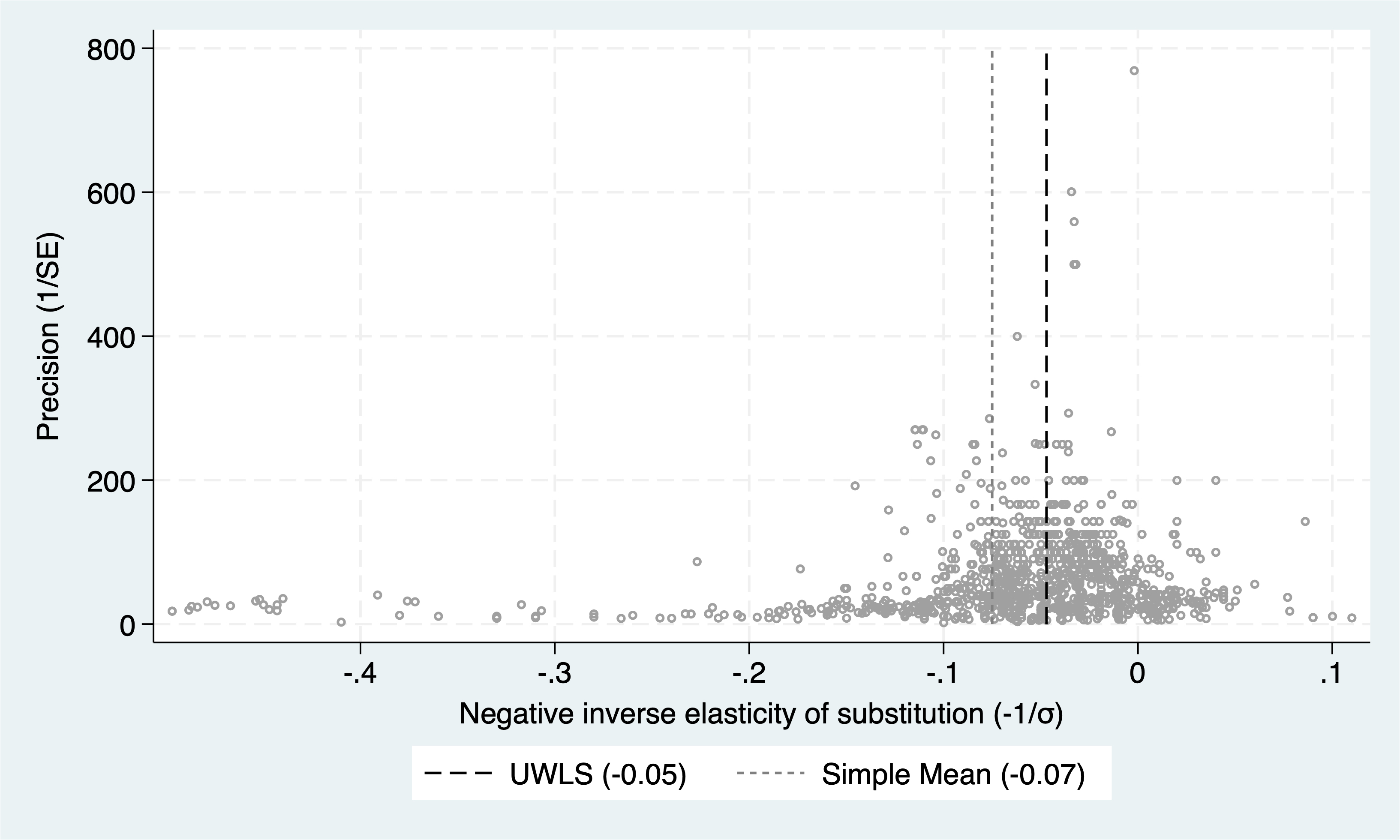}
\begin{tabular*}{\linewidth}{@{}p{\linewidth}@{}}
\scriptsize \emph{Notes:} The vertical lines represent the unweighted arithmetic mean of the estimates ($-0.0749$) and the precision-weighted average ($-0.0470$), calculated via Unrestricted Weighted Least Squares (UWLS). SE denotes the standard error. For visual clarity, the plot excludes 26 outlying estimates with values below $-0.5$.
\end{tabular*}
\end{figure}

\autoref{fig:funnel} reveals noticeable asymmetry, with a longer left tail suggesting that studies reporting more negative coefficients are overrepresented. The gap between the simple mean of the coefficients ($-0.075,$ implying $\sigma \approx 13.3$) and the Unrestricted Weighted Least Squares (UWLS) coefficient ($-0.047$, implying $\sigma \approx 21.3$) is indicative of selective reporting. Following \citet{stanley2023unrestricted}, we use the UWLS as a robust benchmark for the precision-weighted average. The precision-weighted average is meaningfully smaller in magnitude than the unweighted average, suggesting that the most negative results are concentrated among the least precise coefficients.  However, because graphical interpretations and simple weighting can only be indicative of the underlying selection process, we employ formal meta-regression techniques to quantify the extent of publication selection bias and recover a bias-corrected coefficient, which serves as the basis for our implied elasticity estimates.

To assess and accommodate this pattern more formally, we perform meta-regression models based on the relationship between reported coefficients and their standard errors \citep{stanley2008meta}. The baseline Funnel Asymmetry Test (FAT) and Precision-Effect Test (PET) are specified as:
\begin{equation} 
\label{FATPET}
\hat{\theta}_{ij} = \beta_0 + \beta_1 \cdot SE_{ij} + \epsilon_{ij}, 
\end{equation}
where $\hat{\theta}_{ij}$ is the $i$-th reported coefficient (the negative inverse elasticity) from study $j$. In this specification, the coefficient on SE, $\beta_1$, measures publication bias, while the intercept, $\beta_0$, represents the "true" effect beyond bias. To address heteroskedasticity, we estimate \autoref{FATPET} using Weighted Least Squares (WLS) with inverse-variance weights ($w_{ij} = 1/SE_{ij}^2$). This is numerically equivalent to regressing the $t$-statistic on precision ($1/SE_{ij}$):
\begin{equation} 
\label{FATPET_trans}
t_{ij} = \beta_1 + \beta_0 (1/SE_{ij}) + \nu_{ij}, 
\end{equation}
where $t_{ij} = \hat{\theta}_{ij}/SE_{ij}$. Following the decision rule in \citet{stanley2014meta}, we also employ the Precision-Effect Estimate with Standard Error (PEESE) model, which assumes that publication bias is proportional to the variance ($SE^2$):
\begin{equation} 
\label{PEESE}
\hat{\theta}_{ij} = \beta_0 + \beta_1 \cdot SE_{ij}^2 + \epsilon_{ij}. 
\end{equation}
When estimated via WLS with $1/SE_{ij}^2$ weights, the PEESE model transforms into a regression of the $t$-statistic on precision and the standard error itself:
\begin{equation} 
\label{PEESE_trans}
t_{ij} = \beta_0 (1/SE_{ij}) + \beta_1 SE_{ij} + \nu_{ij}.
\end{equation}
This quadratic specification often provides a more accurate correction for the magnitude of the effect when a non-zero effect exists.

To ensure our results are not driven by specific weighting assumptions or study-level heterogeneity, we implement two additional linear specifications. First, we use study-level weights ($1/(SE_{ij} \cdot k_j)$), which balance the influence of individual papers while simultaneously accounting for the precision of each estimate. This prevents studies with a high volume of reported findings from dominating the results. Second, we estimate a Fixed Effects (FE) model to account for study-specific unobserved heterogeneity. By focusing on the within-study variation between reported coefficients and their standard errors, the FE specification provides a robust check against the possibility that our results are driven by unobserved study-level heterogeneity or the specific reporting idiosyncrasies of certain research groups. We do not pursue an instrumental-variable version of these meta-regressions, instrumenting the standard error with the inverse square root of the sample size, because the two are essentially uncorrelated in our sample and $1/\sqrt{N}$ would be a weak instrument, making such estimates unreliable \citep{irsova2025spurious}.

Beyond these linear approaches, we employ several advanced estimators, including the Weighted Average of Adequately Powered Estimates (WAAP) proposed by \cite{Ioannidis2017}, which reduces bias by focusing on coefficients with sufficient statistical power; the Stem-Based Method of \cite{furukawa2020publication}, a non-parametric approach that builds the meta-analytic estimate from the most precise estimates, selected by a data-driven rule; the Endogenous Kink (EK) model of \cite{bom2019kinked}, which allows publication selection to vary with precision through a piecewise-linear meta-regression; and the Selection Model of \cite{andrews2019identification}, which models the probability of publication as a function of the statistical significance of the results.

\autoref{tab:tab2} presents the results from the standard meta-regression estimators (Panel A) and alternative selection-robust models (Panel B). With the exception of the Stem Method, all specifications yield a statistically significant effect beyond bias, and the overall pattern is consistent with selective reporting linked to estimate precision. The PET coefficient in Column 1 is statistically significant ($-0.038$, implying $\sigma \approx 26$), consistent with a non-zero effect beyond publication selection. Following the decision rule in \citet{stanley2014meta}, we focus on the PEESE coefficient ($-0.046$) for magnitude, which suggests an implied elasticity ($\sigma$) of approximately 22. The study-level weighted specification yields a somewhat higher elasticity ($\sigma \approx 39$), likely because it places additional weight on the most precise estimates while preventing papers with many reported results from dominating the sample. The advanced EK model yields a coefficient identical to the linear PET, and the Selection Model implies a similar elasticity of 22.5. 

To interpret these magnitudes, consider that a higher elasticity implies a weaker relative-wage response to immigration. For instance, an elasticity of 13 (the uncorrected mean) implies that a 10\% increase in immigrant labor relative to native labor lowers immigrant wages relative to native wages by about 0.77\%, whereas an elasticity of 22 (corrected for publication bias) reduces that response to about 0.45\%.\footnote{This is based on the standard relationship $\Delta (w_i/w_n) / (w_i/w_n) = -1/\sigma \cdot \Delta (L_i/L_n) / (L_i/L_n)$.} Thus, moving from an elasticity of 13 to 22 reflects an approximately 40\% reduction in the estimated relative-wage pressure from immigration.\footnote{Difference in impact: $0.77\% - 0.45\% = 0.32\%$. Relative reduction: $0.32\%/0.77\% \approx 0.42$.} This pattern reflects the distinction between two different types of labor market adjustments. When the elasticity of substitution is high, it means that immigrants and natives are close substitutes, i.e., they perform similar tasks and can readily replace each other in production. In such cases, employers can substitute between the two groups with little friction, so a given shift in relative labor supply translates into only a small movement in relative wages.

\begin{singlespace}
\begin{table}[H]\scriptsize
\centering
\caption{MRA and advanced estimators suggest significant publication selection bias\label{tab:tab2}}
\begin{tabular*}{\linewidth}{@{\extracolsep{\fill}}lcccc@{}}
\toprule
\textbf{Panel A: MRA Estimators} & (1) & (2) & (3) & (4) \\
                     & PET & PEESE & Study-level & Fixed Effects \\
\midrule
Publication bias & -1.118\sym{*} & -11.244\sym{*} & -1.512\sym{**} & -0.968\sym{**} \\
& (0.572) & (5.762) & (0.641) & (0.474) \\
\addlinespace
Effect beyond bias & -0.038\sym{***} & -0.046\sym{***} & -0.026\sym{***} & -0.041\sym{***} \\
 & (0.011) & (0.009) & (0.008) & (0.007) \\
\addlinespace
\textit{Implied Elasticity ($\sigma$) }& 26.1 & 21.9 & 39.0 & 24.6 \\
\midrule
\textbf{Panel B: Advanced Estimators} & (5) & (6) & (7) & (8) \\
                     & WAAP & Stem Method & EK & Selection Model \\
\midrule
Publication bias       & --- & --- & -1.118\sym{***} & P = 0.760 \\
& & & (0.165) & (0.068) \\
\addlinespace
Effect beyond bias & -0.045\sym{***} & -0.027 & -0.038\sym{***} & -0.044\sym{***} \\
& (0.002) & (0.021) & (0.002) & (0.002) \\
\addlinespace
\textit{Implied Elasticity ($\sigma$)} & 22.0 & 37.3 & 26.1 & 22.5 \\
\bottomrule 
\end{tabular*}
\begin{tabular*}{\linewidth}{@{}p{\linewidth}@{}}
\scriptsize \emph{Notes:} Column (1) presents the FAT-PET test using a $t$-statistic transformation, which is numerically equivalent to WLS with inverse-variance weights ($1/SE_{ij}^2$) as shown in \autoref{FATPET_trans}. Column (2) reports PEESE results as specified in \autoref{PEESE}, also estimated via WLS with $1/SE_{ij}^2$ weights, which is equivalent to \autoref{PEESE_trans}. PEESE is reported here because the PET intercept is statistically significant \citep{stanley2014meta}. Column (3) is based on the specification in \autoref{FATPET} and is weighted by $1/(SE_{ij} \cdot k_j)$, where $k_j$ is the number of estimates per study. Column (4) applies a study-level fixed-effects estimator to the relationship in \autoref{FATPET_trans} to account for unobserved study-level heterogeneity. Column (5) reports Weighted Average of Adequately Powered Estimates \citep{Ioannidis2017}. Column (6) follows Stem-based method \citep{furukawa2020publication}. Column (7) uses Endogenous Kink piecewise approach \citep{bom2019kinked}. Column (8) presents Selection Model \citep{andrews2019identification}, where $P$ denotes the relative publication probability of results that are not statistically significant at the 5\% level. In Panel A (Columns 1--4), standard errors are clustered at the study level. In Panel B, the parenthetical values are the standard errors produced by each estimator's own procedure (WAAP, the Stem method, Endogenous Kink, and the Selection Model) and are not study-clustered. The implied elasticity ($\sigma$) is calculated as $\sigma = -1/(\text{Effect beyond bias})$ using the estimates prior to rounding. \sym{*} \(p<0.10\), \sym{**} \(p<0.05\), \sym{***} \(p<0.01\).

\end{tabular*}
\end{table}
\end{singlespace}

However, the limited relative-wage response does not imply no impact: part of the burden may fall on the extensive margin. Some natives may exit employment or be displaced to other sectors. In other words, displacement may become more important than wage compression. This distinction is emphasized by \cite{borjas2003labor}, who notes that average wage effects may understate the true economic pressure from immigration if substitution is high. Similarly, \cite{smith2003labour} explains that partial equilibrium models often show wage declines, while general equilibrium frameworks highlight the role of reallocation and sectoral mobility.

\citet{ottaviano2012rethinking} argue that immigration's wage impact depends critically on the degree of substitutability. In contexts where immigrants and natives are not perfect substitutes, the effects are more diffuse and may involve complementarities in production. But when substitutability is high, as our corrected elasticity of 22 suggests, wages become less responsive, while employment composition becomes the primary margin of adjustment. Therefore, the finding that a 10\% increase in relative immigrant labor lowers the immigrant-to-native relative wage by only 0.45\% when the elasticity is 22 does not imply an absence of labor market competition. Rather, it suggests that labor markets adjust via both wages and employment, with the wage margin being muted under high substitutability.

\section{Heterogeneity}
\label{sec:het}

While publication bias distorts meta-analytic results by favoring significant results, heterogeneity reflects genuine differences across studies due to variation in data, methodology, and context. Identifying the key drivers of heterogeneity is critical for understanding when and why studies report higher or lower elasticities of substitution between immigrant and native labor \citep{havranek2011vertical}. In this section, we explore the extent and sources of heterogeneity using Bayesian Model Averaging (BMA) and Frequentist Model Averaging (FMA).

The literature on the elasticity of substitution spans multiple countries, time periods, data types, and estimation strategies. As such, reported coefficients are likely to reflect underlying economic relationships as well as structural and methodological variation. We categorize sources of heterogeneity into four domains: data characteristics, structural variation, estimation characteristics, and publication characteristics. \autoref{tab:Het_vars} summarizes the variables used in our heterogeneity models. For each, we coded 1,087 observations from 40 studies.\footnote{The number of observations differs slightly from the publication bias analysis. We excluded 4 observations (1 study) because they were estimated using a multi-country sample, which precluded the assignment of country-specific moderator variables, such as the immigration share.}

%%vars used
\begin{singlespace}
\begin{table}[H]\scriptsize
\begin{center}
\caption{Characteristics used to explain heterogeneity\label{tab:Het_vars}}
\begin{tabular*}{\textwidth}{@{\hskip\tabcolsep\extracolsep\fill}p{0.3\textwidth}p{0.7\textwidth}@{}}
\toprule
\textbf{Category}           & \textbf{Variables} \\
\midrule
Data characteristics        & Number of cells, Annual frequency data \\
Structural variation        & All workers, High level of experience, Low level of experience, High \newline level of education, Low level of education, Top 6 languages, Male, \newline Female, Immigrant population \\
Estimation characteristics          & Log of mean wages, Annual wage, Monthly wage, Daily wage, Hourly wage, National, OLS \\
    & Time fixed effects, Person fixed effects, Skill fixed effects \\
Publication characteristics & Impact factor, Citations, Published, Top 5 journals, \newline Publication year \\
\bottomrule 
\end{tabular*}
\begin{tabular*}{\textwidth}{@{\hskip\tabcolsep\extracolsep\fill}cc}
\multicolumn{2}{p{\textwidth}}{\scriptsize \emph{Notes:} Details on each variable, including definition and summary statistics, are available in \autoref{tab:descr}.}
\end{tabular*}
\end{center}
\vspace{-0.4cm}
\end{table}
\end{singlespace}

\begin{figure}[!htbp]
\centering
\caption{Model inclusion in Bayesian model averaging}
\includegraphics[width=1\linewidth]{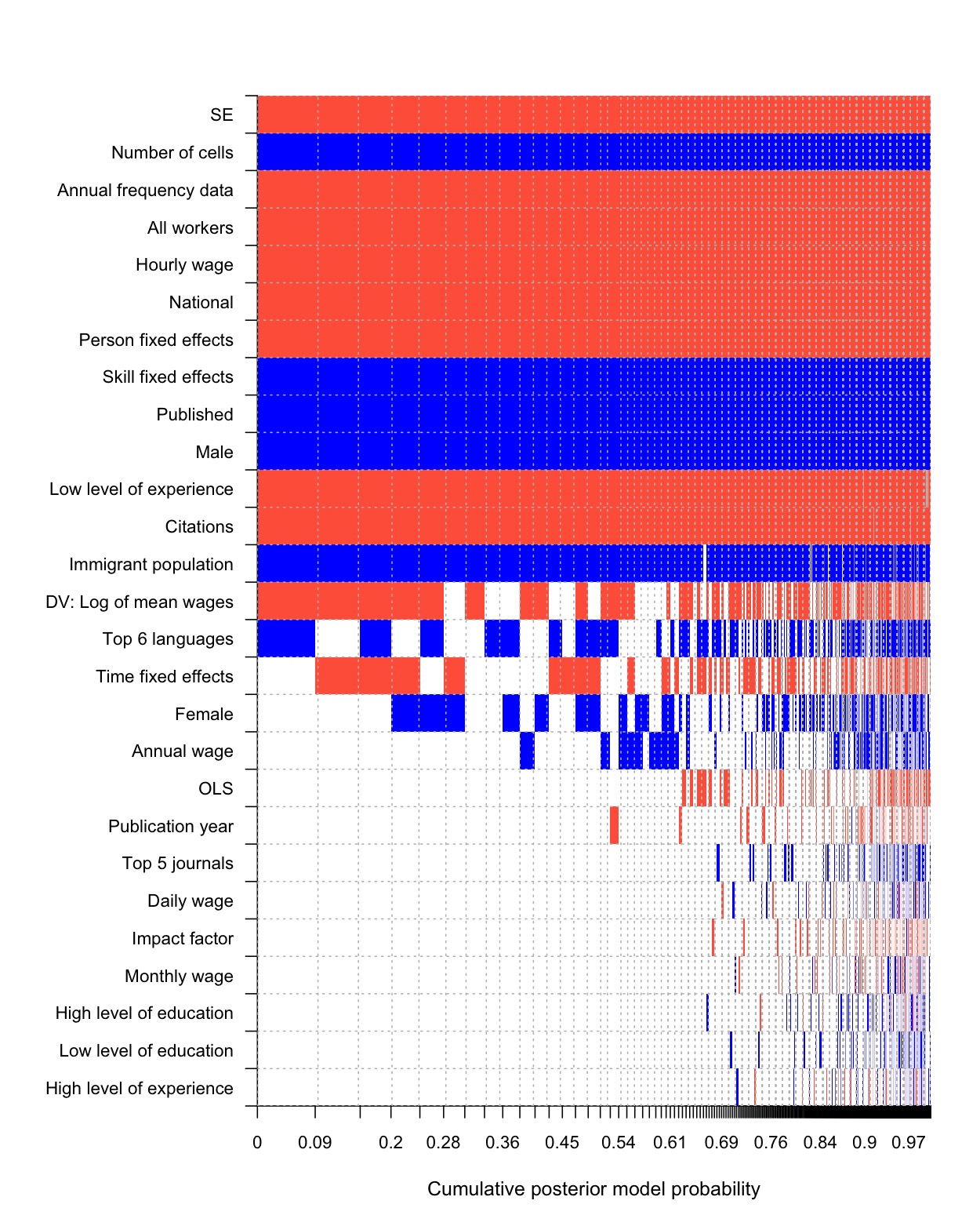}
\label{fig:BMA}
\begin{tabular*}{1\hsize}{@{\hskip\tabcolsep\extracolsep\fill}cccc}
\multicolumn{4}{p{1\hsize}}{\scriptsize \emph{Notes:} The response variable is the reported coefficient ($-1/\sigma$). Columns denote individual models; variables are sorted by posterior inclusion probability in descending order. The horizontal axis denotes cumulative posterior model probabilities. Blue color (positive sign) indicates that the variable is associated with a less negative coefficient, implying a higher elasticity of substitution. Red color (negative sign) indicates a more negative coefficient, implying a lower elasticity. No color = the variable is not included in the model. Numerical results of the BMA exercise are reported in \autoref{tab:BMA}, detailed description of all variables is available in \autoref{tab:descr}.}
\end{tabular*}
\end{figure}

To account for model uncertainty, we estimate both BMA and FMA models, using the coefficient (i.e., $\theta = -1/\sigma$) as the dependent variable. We implement BMA using the Unit Information Prior (UIP) for the $g$-prior (UIP centers the $g$-prior around zero with a variance equal to the number of observations) and a uniform prior over model space as the baseline specification. This combination follows common practice in economic meta-analyses \citep{Eicher2011,havranek2020reporting}. It is parsimonious because the baseline prior is simple and transparent, yet flexible because BMA still compares many alternative combinations of moderators and weights them by the data. To ensure robustness, we perform a series of additional estimations using alternative $g$-prior settings, namely the benchmark BRIC prior \citep{fernandez2001benchmark}, the empirical Bayes local (EBL) prior \citep{feldkircher2009benchmark}, and the hyper-$g$ prior \citep{liang2008mixtures}, combined with both uniform and random model priors. The results of these robustness checks are consistent with the baseline model and are presented in Appendix~\ref{sec:apphet} (specifically in \autoref{tab:bma_robust_random} and \autoref{tab:bma_robust_uniform}, and visually confirmed in Figures~\ref{fig:BMA_UIP_random}--\ref{fig:PMP_BRIC_uniform}).\footnote{The large magnitude of the Constant in some specifications is a mechanical result of the inclusion of non-centered logarithmic variables (such as the logarithm of publication year) and does not affect the interpretation of the posterior means of the other moderators.} All BMA specifications were estimated using Markov Chain Monte Carlo (MCMC) sampling with 3 million iterations (after a 1 million burn-in), and the full model space was explored up to 10,000 models. \autoref{fig:BMA} shows posterior inclusion probabilities (PIPs) across all variables, while \autoref{tab:BMA} reports corresponding coefficient estimates.

%BMA FMA
\begin{singlespace}
\begin{table}[!btph]\scriptsize
\begin{center}
\caption{Why do estimates of the elasticity of substitution differ?\label{tab:BMA}}
\begin{tabular*}{1\hsize}{@{\hskip\tabcolsep\extracolsep\fill}l*{7}{c}@{}}
\toprule
                      & \multicolumn{3}{c}{Bayesian model averaging}        & \multicolumn{3}{c}{Frequentist check}  \\ \cline{2-4} \cline{5-7}
                      & PIP    & Post. mean & Post. SD & Coef.  & SE    & \textit{p}-value \\
                                            
 \midrule
Constant                           & 1.00 & 0.64       & NA      & 11.00       & 11.95 & 0.36 \\
SE                                & 1.00 & -0.52      & 0.08    & -0.54       & 0.08  & 0.00 \\
\midrule
\multicolumn{3}{l}{\textbf{Data characteristics}}                   &       &        &       &         \\
Number of cells                    & 1.00 & 0.06      & 0.01    & 0.07        & 0.01  & 0.00 \\
Annual frequency data              & 1.00 & -0.08     & 0.01    & -0.07       & 0.01  & 0.00 \\
\midrule
\multicolumn{3}{l}{\textbf{Structural variation}}                   &       &        &       &         \\
All workers                        & 1.00 & -0.05     & 0.01    & -0.05       & 0.01  & 0.00 \\
High level of experience          & 0.03 & 0.00      & 0.00    & 0.01        & 0.01  & 0.55 \\
Low level of experience            & 1.00 & -0.05     & 0.01    & -0.04       & 0.01  & 0.00 \\
High level of education           & 0.04 & 0.00      & 0.00    & 0.01        & 0.01  & 0.47 \\
Low level of education             & 0.03 & 0.00      & 0.00    & 0.01        & 0.01  & 0.53 \\
Top 6 languages                     & 0.54 & 0.02      & 0.02    & 0.04        & 0.02  & 0.02 \\
Male                               & 1.00 & 0.04      & 0.01    & 0.04        & 0.01  & 0.00 \\
Female                        & 0.39 & 0.01      & 0.01    & 0.02        & 0.01  & 0.03 \\
Immigrant population                      & 0.98 & 0.34      & 0.10    & 0.25        & 0.09  & 0.01 \\
\midrule
\multicolumn{3}{l}{\textbf{Estimation characteristics}}             &       &        &       &         \\
DV: Log of mean wages   & 0.64 & -0.01     & 0.01    & -0.02       & 0.01  & 0.07 \\
Annual wage            & 0.21 & 0.01      & 0.01    & 0.02        & 0.01  & 0.23 \\
Monthly wage            & 0.04 & 0.00      & 0.00    & 0.01        & 0.02  & 0.65 \\
Daily wage              & 0.04 & 0.00      & 0.00    & 0.02        & 0.02  & 0.39 \\
Hourly wage             & 1.00 & -0.06     & 0.01    & -0.06       & 0.01  & 0.00 \\
National                           & 1.00 & -0.06     & 0.01    & -0.05       & 0.01  & 0.00 \\
OLS                                & 0.11 & 0.00      & 0.00    & -0.01       & 0.01  & 0.07 \\
Time fixed effects                           & 0.47 & -0.01     & 0.01    & -0.01       & 0.01  & 0.20 \\
Person fixed effects                        & 1.00 & -0.08     & 0.01    & -0.09       & 0.01  & 0.00 \\
Skill fixed effects                             & 1.00 & 0.07      & 0.01    & 0.07        & 0.01  & 0.00 \\
\midrule
\multicolumn{3}{l}{\textbf{Publication characteristics}}            &       &        &       &         \\
Impact factor                       & 0.04 & 0.00      & 0.00    & -0.01       & 0.01  & 0.35 \\
Citations                          & 1.00 & -0.02     & 0.01    & -0.02       & 0.01  & 0.00 \\
Published                    & 1.00 & 0.05      & 0.01    & 0.06        & 0.01  & 0.00 \\
Top 5 journals                       & 0.05 & 0.00      & 0.01    & 0.03        & 0.02  & 0.14 \\
Publication year & 0.06 & -0.25     & 1.32    & -3.39       & 3.62  & 0.35 \\
\midrule
Studies                &      & 40        &         &             & 40    &      \\
Observations           &      & 1,087     &         &             & 1,087 &      \\
\bottomrule 
\end{tabular*}
\begin{tabular*}{1\hsize}{@{\hskip\tabcolsep\extracolsep\fill}cccc}
\multicolumn{4}{p{1\hsize}}{\scriptsize \emph{Notes:} PIP = posterior inclusion probability. For the interpretation of the posterior inclusion probability, we follow the guidelines offered by \cite{kass1995bayes} to be able to evaluate the importance of each explanatory variable: $0.5 <PIP <0.75$: weak effect, $0.75 <PIP <0.95$: substantial effect, $0.95 <PIP <0.99$: strong effect, $0.99 <PIP$: decisive effect. SD = standard deviation. The table shows unconditional moments for BMA. \textit{Number of cells}, \textit{citations}, and \textit{publication year} enter in logarithms. Detailed description of all variables is available in \autoref{tab:descr}.}
\end{tabular*}
\end{center}
\vspace{-0.4cm}
\end{table}
\end{singlespace}

First, we clarify how to interpret the dependent variable used in our analysis. All estimates in our dataset are coded as the negative inverse of the elasticity of substitution ($\theta = -1/\sigma$). As a result, a more negative value of $\theta$ indicates a lower elasticity ($\sigma$), meaning weaker substitutability between native and immigrant labor. Conversely, values closer to zero (less negative) imply higher elasticity and thus greater substitutability. Several variables emerge as robust predictors of the reported coefficients. 

Larger standard errors are consistently associated with more negative coefficients. Following \citet{stanley2014meta}, this correlation suggests that as precision decreases, reported results tend to be more extreme. The correlation between the coefficient and the inverse square root of sample size is negligible in our data ($r=0.016$), whereas the correlation with the standard error is notably stronger ($r=-0.24$). This suggests that the asymmetry is not driven simply by smaller sample sizes but is more consistent with selective reporting among less precise estimates. 

The granularity of the skill-experience-region-gender-time matrix, measured by the logarithm of the number of cells, is a decisive predictor. As the number of cells increases, coefficients become less negative, implying higher elasticity. Immigrants and natives appear more substitutable within precisely defined niches. Conversely, studies using samples of all workers (full-time as well as part-time workers) report more negative coefficients, implying lower elasticity. 

Methodological choices also shift results. Studies using annual frequency data report more negative coefficients (i.e., lower $\sigma$). This aligns with the labor economics distinction between short-run frictions and long-run adjustments. Annual data likely captures nominal rigidities or slow native mobility that result in lower implied elasticities compared to the longer-term adjustments captured in decennial census data \citep{borjas2014immigration}. The use of hourly wages as the dependent variable is also associated with more negative coefficients (lower $\sigma$). This is in line with the view that hourly wages are closer to the wage measure emphasized in structural models than annual or monthly earnings, which can embed labor-supply adjustments \citep{Borjasetal2008}. Similarly, the national-level approach yields more negative coefficients. This result accords with the evidence that regional estimates often conflate labor supply shocks with compensating native out-migration. When natives relocate in response to immigration, local wage impacts are dissipated, potentially biasing local estimates toward zero (implying near-infinite substitutability). By contrast, national-level analyses internalize these geographic shifts within the aggregate labor market, providing more conservative and theoretically robust estimates of substitutability \citep{jaeger2018shift}.

Estimates focusing on low-experience workers are systematically more negative, indicating lower substitutability. This aligns with the theory that entry-level immigrants and natives may possess more distinct, non-transferable skill sets compared to more experienced workers who have assimilated into general labor market tasks \citep{borjas2003labor,borjas2012com_est}. Conversely, male samples are associated with coefficients closer to zero, indicating higher substitutability. A similar increase in elasticity ($\sigma$) is observed in markets with a higher immigrant population share. This pattern is consistent with labor-market adaptation to immigration: firms adjust production processes and native workers specialize in tasks where they retain a comparative advantage \citep{peri2012effect,ottaviano2012rethinking}. Such adjustment dampens the relative-wage response to immigration-induced supply shocks, which appears in the estimating equation as a higher effective elasticity.

The inclusion of person fixed effects, i.e., fixed effects for individual characteristics such as age, location, and gender, results in more negative values (lower $\sigma$). As suggested by \citet{Peri2009}, absorbing these characteristics helps uncover persistent native-immigrant differences in task specialization and imperfectly transferable skills that are otherwise masked in aggregate data. In contrast, including skill fixed effects (education and experience) shifts coefficients toward zero, suggesting that a portion of the "imperfect substitution" observed in simpler models may be a byproduct of unobserved skill-level variation rather than a fundamental trait of the labor market.

Published studies report coefficients closer to zero (higher $\sigma$), whereas more highly cited studies report more negative values of the coefficient (lower $\sigma$). As argued by \citet{peri2012effect}, findings of low substitutability are often framed as more impactful because they allow for richer, heterogeneity-based frameworks that explain how immigration can influence productivity and task specialization. Alternatively, as \citet{borjas2012com_est} emphasizes, identifying a low degree of substitution could be positioned as a more complex and critical departure from textbook assumptions, essential for understanding the potential for native wage depression. 

Finally, the choice of dependent variable transformation (the log of mean wages versus the mean of log wages) receives a weak PIP = 0.64. While this technical choice has been the subject of significant debate in primary studies, our results suggest it is a less robust driver of heterogeneity than structural or reporting characteristics.

\subsection{Best-Practice Estimate}

To derive a consensus estimate of the native-immigrant substitution elasticity, we calculate best-practice estimates, in the tradition of \citet{havranek2015measuring}, by substituting sensible values into our preferred MRA model. We use a model containing moderators with a Posterior Inclusion Probability (PIP) above 0.60 with the exception of the female dummy, which is retained so that the baseline reflects the average gender composition of the literature. The model is estimated by ordinary least squares with standard errors clustered at the study level; the exact specification is available in the replication package. 

To minimize researcher degrees of freedom, we define "best practice" as a published study employing the most granular data available (maximum number of cells, annual frequency data, and hourly wages) and a set of fixed effects (person and skill). We net out the effect of publication selection bias by setting the standard error to zero. For variables where a single "best" choice is not theoretically dominant, we use sample means to ensure a consensus view.  Specifically, we set the gender dummies to their sample averages (the values can be found in \autoref{tab:descr}) to account for the fact that half of the literature examines both genders combined, while the rest focuses on specific subgroups (male and female). Similarly, the low-experience dummy and the share of immigrants are set to their respective sample means.  In contrast, for variables associated with academic impact and peer recognition, we focus only on published studies and set the citation count (normalized as the logarithm of citations per year) to its sample maximum. This ensures that the best-practice estimate prioritizes evidence from the most influential research. 

\begin{table}[htbp]\scriptsize
\centering
\caption{Best-practice estimates of the native-immigrant substitution elasticity}
\label{tab:BPE}
\begin{tabular*}{1\hsize}{@{\hskip\tabcolsep\extracolsep\fill}l*{4}{c}@{}}
\toprule
Specification & Estimate ($-1/\sigma$) & 95\% CI & Implied $\sigma$ \\
\midrule
\textit{    Regional} & & & \\
\textbf{Mean of log wages} &\textbf{-0.06} & \textbf{[-0.11, -0.00]} & \textbf{16.9} \\
Log of mean wages & -0.08 & [-0.14, -0.02] & 12.3 \\
\midrule
\textit{    National} & & & \\
Mean of log wages  & -0.12 & [-0.21, -0.04] & 8.2 \\
Log of mean wages  & -0.14 & [-0.22, -0.07] & 6.9 \\
\midrule
\textit{    Regional + mean of log wages} & & & \\
Full-time only & -0.03 & [-0.09, 0.02] & 29.4 \\
All workers (incl. part-time) & -0.08 & [-0.14, -0.02] & 12.5 \\
\bottomrule
\end{tabular*}
\begin{tabular*}{1\hsize}{@{\hskip\tabcolsep\extracolsep\fill}cccc}
\multicolumn{4}{p{1\hsize}}{\scriptsize \emph{Notes:} Details on each variable, including definitions and summary statistics, are available in \autoref{tab:descr}. The baseline estimate (in bold) evaluates worker and market characteristics at their sample means: gender ($\textit{male}$ at 0.41, $\textit{female}$ at 0.09), experience ($\textit{low level of experience}$ at 0.07), and immigrant share ($\textit{immigrant population}$ at 0.09). To represent best practice, we condition on the estimate coming from a published study (dummy set to 1), assume maximum data granularity ($\textit{log number of cells}$ set to 3.9, with dummies for annual data and hourly wages set to 1), and set citation impact to its sample maximum ($\textit{log citations per year}$ set to 2.7). The baseline holds the worker sample composition at its mean ($\textit{all workers}$ at 0.55), while the bottom panel explicitly toggles this condition. We also assume the underlying primary estimates control for person and skill fixed effects (respective dummies set to 1). Standard errors are clustered at the study level.}
\end{tabular*}
\end{table}

The results are summarized in \autoref{tab:BPE}. Our best-practice estimates reflect the highest levels of data granularity (annual frequency and hourly wages). As shown in our previous analysis, these settings systematically yield more conservative elasticity estimates than a simple correction for publication bias would suggest. By conditioning on study-design features associated with more detailed labor-market measurement, the best-practice estimate provides a middle ground: it corrects for the downward bias of the raw sample mean (13.3) but avoids the potentially inflated estimates (which generally exceed 22) found when correcting for bias without conditioning on study-design features. In our baseline regional specification (shown in bold), using the mean of log wages and representing the average labor force composition, the predicted coefficient is $-0.06$, which yields an implied elasticity ($\sigma$) of approximately 17. When toggling the geographic scale to a national level, the coefficient shifts to $-0.12$ ($\sigma \approx 8$), confirming that geographic scale is one of the primary drivers of the variation in the literature. Consistent with concerns raised by \citet{borjas2012com_est} about Jensen's inequality and the construction of cell-level wages, using the log of mean wages consistently yields more negative estimates (lower $\sigma$) than the mean of log wages baseline.

Finally, we perform a sensitivity analysis regarding the labor force definition. Our results show that studies focusing exclusively on full-time workers yield higher elasticity ($\sigma \approx 29$), while those encompassing the entire labor force (including part-time workers) suggest lower substitutability ($\sigma \approx 13$). By using this parsimonious approach and sample-wide means for demographic variables, we provide an assessment of the elasticity that is less sensitive to subjective judgments across less relevant moderators.

\section{Conclusion}
\label{sec:concl}
This meta-analysis offers a comprehensive reassessment of the substitutability between native and immigrant labor. Drawing on 1,091 estimates from 41 studies, we show that the existing literature is shaped by two distinct forces. First, we find that publication selection bias systematically understates the elasticity of substitution. After correcting for selective reporting, the implied elasticity rises from a simple mean of 13 to approximately 22. Second, we find that this correction is only part of the story, as research design choices significantly influence the magnitude of reported effects. Our best-practice estimate, which uses our preferred design-adjusted specification (using hourly wages and annual data frequency) while netting out publication bias, suggests a more conservative consensus of approximately 8 (national) to 17 (regional). The spread between the national and regional estimates confirms that the definition of the labor market, whether localized or aggregate, significantly alters the measured degree of competition between natives and immigrants.

Our findings offer a resolution to the long-standing methodological debate over wage transformations. While \citet{borjas2012com_est} and \citet{ottaviano2012rethinking} disagreed on the validity of the log of mean wages versus the mean of log wages, our analysis suggests that this choice is a significant, albeit secondary, driver of heterogeneity. Once we control for publication bias and modeling scale, the wage definition remains a consistent predictor of reported elasticities, shifting the implied elasticity ($\sigma$) by roughly 1 to 5 units in our best-practice specifications. This confirms that the wage transformation is a systematic factor that researchers should calibrate alongside other data features, not estimation noise.

The policy implications of these results are nuanced. In a theoretical framework, a higher elasticity of substitution implies that relative wages are less sensitive to immigration-induced shifts in relative labor supply, as the shock is spread across a more homogeneous labor pool. While our correction for publication bias alone suggests relatively high substitutability ($\sigma \approx 22$), focusing on the most granular data at the regional level yields a lower consensus estimate in the range of 12 to 17, depending on the wage definition (\autoref{tab:BPE}). This lower elasticity, obtained by conditioning on the most granular measurement choices in the literature, indicates that natives and immigrants possess more distinct, complementary skill sets than the bias correction alone suggests. 

From a methodological standpoint, our findings underscore the need to model bias and heterogeneity jointly. The convergence of linear, non-linear, advanced, and model-averaging approaches enhances the credibility of our estimates and provides a clearer benchmark for future research. While limitations regarding country-specific institutional differences remain, this study provides the first systematic synthesis of the structural elasticity of substitution. By correcting for selective reporting and identifying the drivers of research variation, we lay a more reliable foundation for understanding how immigration reshapes modern labor markets.

\section*{Use of Generative AI}
\addcontentsline{toc}{section}{Use of Generative AI}

The authors used generative AI tools, including Claude (Anthropic) and ChatGPT (OpenAI), during
manuscript preparation to improve language, readability, and formatting, to cross-check the accuracy
of citations and the internal consistency of the text and reported numbers, and to edit and
consistency-check the replication package. AI tools were not used for study selection, data
collection, data coding, estimation choices, analytical decisions, interpretation of results, or the
formulation of conclusions. All estimates were collected by hand from the primary studies; AI
assistance with code was limited to editing, consistency checking, and formatting, and all analyses
were run and verified by the authors. Every result reported in the paper is fully reproducible
without AI tools using the replication package available at
\url{https://meta-analysis.cz/migrant/}. The authors reviewed and verified all AI-assisted output
and take full responsibility for the content of the manuscript.

\newpage

\onehalfspacing
\phantomsection
\addcontentsline{toc}{section}{References}
\setlength{\bibsep}{4pt} %compact bibliography
\begin{footnotesize}
\begin{multicols}{2}
\bibliographystyle{stylebib}

\bibliography{immi_wp}
\end{multicols}
\end{footnotesize}
\onehalfspacing
%%\doublespacing
\normalsize

\appendix
\clearpage

% Add "Appendix" as a title
\section*{Appendix}
%\addcontentsline{toc}{section}{Appendix} % Add to Table of Contents (optional)

% Redefine section numbering to use letters (A, B, C, ...)
\renewcommand\thesection{\Alph{section}}

% Redefine table and figure numbering for appendices
\renewcommand\thetable{\thesection\arabic{table}}
\renewcommand\thefigure{\thesection\arabic{figure}}

% Start Appendix A
\section{Data}
\label{app:Data}

% Reset counters for tables and figures in each appendix section
\setcounter{table}{0}
\setcounter{figure}{0}

\begin{table}[!htbp] \centering \footnotesize 
		\caption{Studies included in the meta-analysis\label{tab:app:studies}}
		\begin{tabular}{lll}
    \toprule
   \cite{akbari2013educational} & \cite{albert2022immigration} & \cite{Aydemir2007} \\
    \cite{boateng2019} & \cite{Borjasetal2008} & \cite{Borjas2010} \\
    \cite{borjas2012com_est} & \cite{bound2006international} & \cite{bratsberg2014immigration} \\
    \cite{brucker2008migration} & \cite{busch2020should} & \cite{Card2009} \\
    \cite{chiswick1985immigrants} & \cite{Damuri2010} & \cite{edo2014immigration} \\
    \cite{edo2015selective} & \cite{edo2017immigration} & \cite{etzo2021complementarities} \\
    \cite{felbermayr2010restrictive} & \cite{gentili2024drives} & \cite{gerfin2010effects} \\
    \cite{goldsmith2020bartik} & \cite{gunadi2019inquiry} & \cite{haas2013employ} \\
    \cite{hill2018noncitizen} & \cite{jaeger1996} & \cite{lebow2024immigration} \\
    \cite{llull2018immigration} & \cite{lu2023detecting} & \cite{Manacorda2012} \\
    \cite{nelson1999earning} & \cite{nguyeneffect} & \cite{OttavPeri2008} \\
    \cite{ottaviano2012rethinking} & \cite{Peri2007} & \cite{raphael2009immigration} \\
    \cite{romiti2011immigrants} & \cite{sharpe2020competes} & \cite{signorelli2024skilled} \\
    \cite{wei2016imperfect} & \cite{wei2019substitution} & \\
    \bottomrule
		\end{tabular} 
\end{table}

\begin{figure}[!htbp]
\centering
\caption{PRISMA diagram}
\includegraphics[width=1\linewidth]{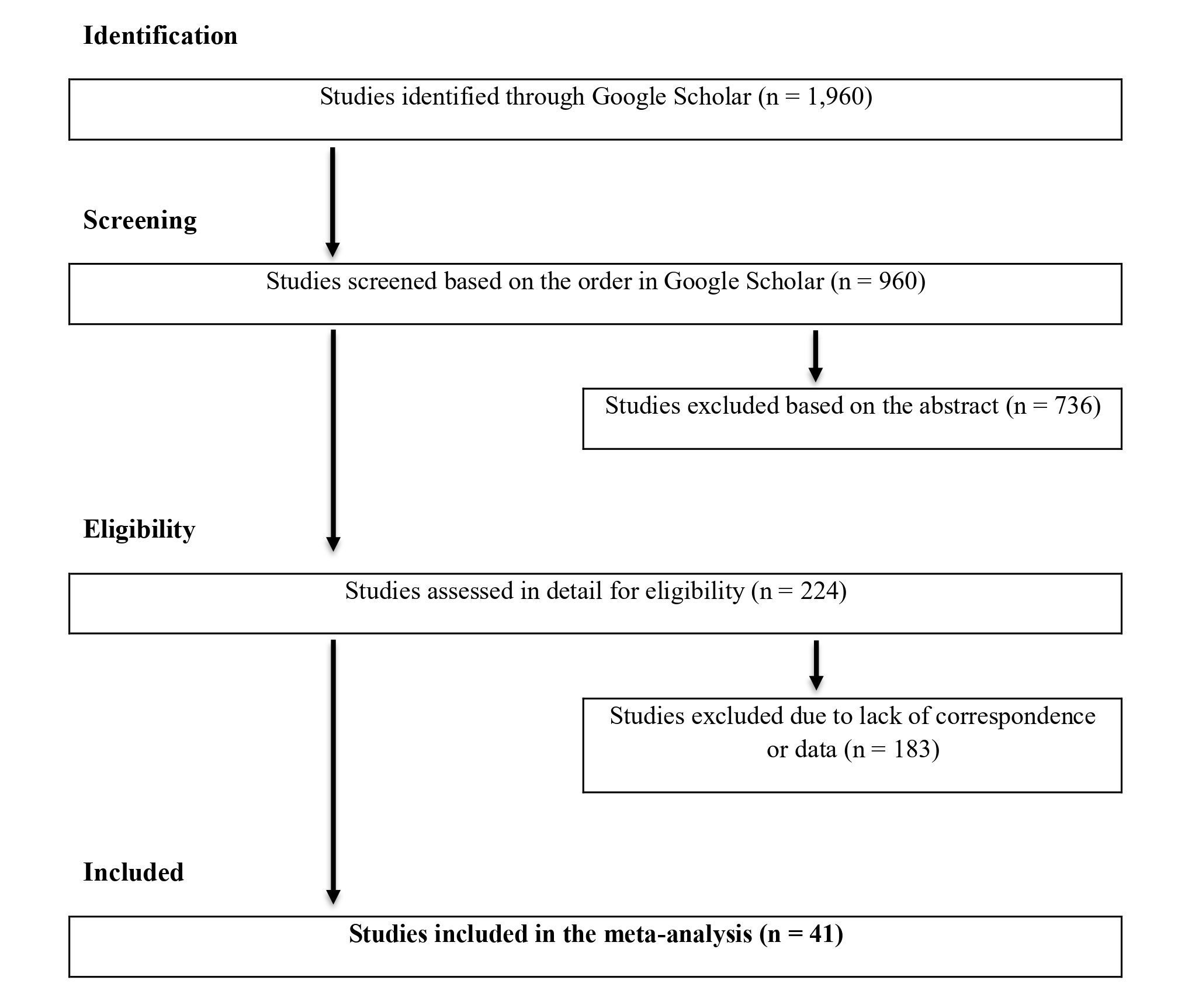}
\label{fig:PRISMA}
\end{figure}

\begin{figure}[!bhtp]
\centering
\caption{Box plot of negative inverse elasticity across countries}
\includegraphics[width=0.8\linewidth]{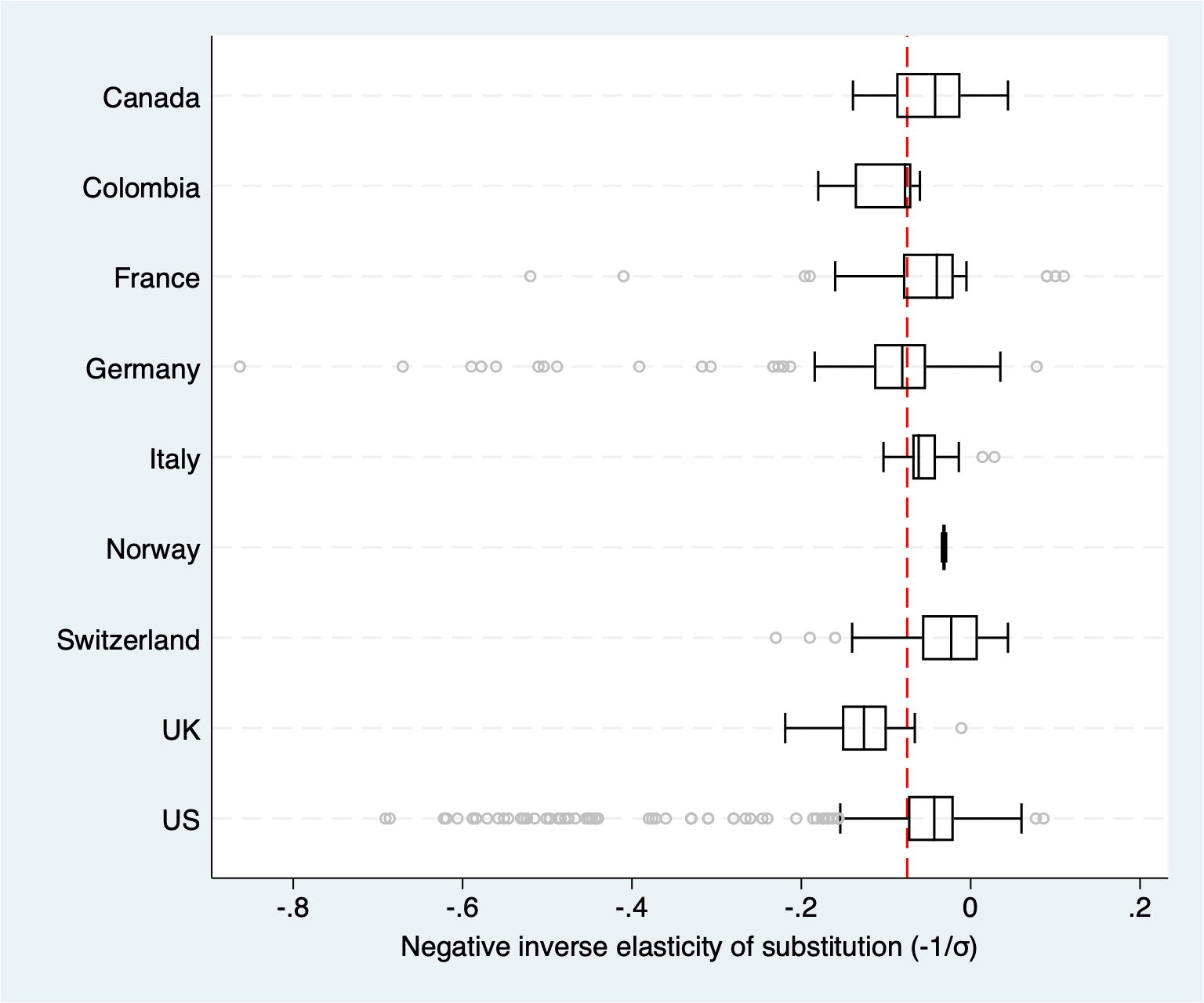}
\label{fig:country}
\begin{tabular*}{1\hsize}{@{\hskip\tabcolsep\extracolsep\fill}cccc}
\multicolumn{4}{p{1\hsize}}{\scriptsize \emph{Notes:} The length of each box represents the interquartile range (P25-P75), and the dividing line inside the box is the median value. The whiskers represent the highest and lowest data points within 1.5 times the range between the upper and lower quartiles. The dots show the outlying estimates with extreme values stacked at the values denoted as "outliers". The red vertical line presents the unweighted mean of all estimates ($-0.075$).}
\end{tabular*}
\end{figure}

\newpage
\noindent \autoref{fig:country} shows a box plot of the coefficient, the negative inverse elasticity ($-1/\sigma$), across countries, revealing that the UK has the least substitutability, while Switzerland has the highest. However, the UK estimate is based on a single study and should be treated with caution. The corresponding summary statistics are quantified in \autoref{tab:tab2_countr}.

%%%%TABLE 2: summary stats countries
\begin{singlespace}
\begin{table}[H]\scriptsize
\begin{center}
\caption{Summary statistics for different countries\label{tab:tab2_countr}}
\begin{tabular*}{\textwidth}{@{\extracolsep{\fill}}p{3cm}cccccccccc}
\toprule
                &         &             & \multicolumn{4}{c}{Unweighted}                           & \multicolumn{4}{c}{Weighted}                             \\\cline{4-7}\cline{8-11}
Country        & Studies & N & Mean & $\sigma$ & \multicolumn{2}{l}{95\% conf. int.} & Mean & $\sigma$ & \multicolumn{2}{l}{95\% conf. int.} \\
\midrule
Australia–Britain–Canada  & 1  & 4   & -0.03  & 28.6  & -0.04 & -0.03 & -0.03 & 28.6  & -0.04 & -0.03 \\
\,\,\,\,\,–Israel–US combined  &   &    &  &  &  & & &   &  &  \\
Canada                          & 3  & 63  & -0.05  & 21.1  & -0.06 & -0.04 & -0.04 & 24.7  & -0.05 & -0.03 \\
Colombia                        & 1  & 30  & -0.10  & 10.3  & -0.11 & -0.08 & -0.10 & 10.3  & -0.11 & -0.08 \\
France                          & 4  & 83  & -0.06  & 17.1  & -0.08 & -0.04 & -0.03 & 33.9  & -0.05 & -0.01 \\
Germany                         & 5  & 136 & -0.12  & 8.5   & -0.14 & -0.09 & -0.09 & 11.3  & -0.10 & -0.07 \\
Italy                           & 2  & 35  & -0.06  & 18.1  & -0.06 & -0.05 & -0.06 & 17.8  & -0.06 & -0.05 \\
Norway                          & 1  & 4   & -0.03  & 31.7  & -0.04 & -0.03 & -0.03 & 31.7  & -0.04 & -0.03 \\
Switzerland                     & 2  & 68  & -0.03  & 34.6  & -0.04 & -0.02 & -0.08 & 12.6  & -0.10 & -0.06 \\
UK                              & 1  & 25  & -0.13  & 7.9   & -0.15 & -0.11 & -0.13 & 7.9   & -0.15 & -0.11 \\
US                              & 21 & 643 & -0.07  & 13.5  & -0.08 & -0.06 & -0.09 & 11.0  & -0.10 & -0.08 \\
US without farmers              & 19 & 609 & -0.05  & 20.3  & -0.05 & -0.04 & -0.05 & 21.4  & -0.05 & -0.04 \\
\cline{2-3}
                                 & 41 & 1,091 &       &       &       &       &       &       &       & \\
\bottomrule 
\end{tabular*}

\vspace{0.2cm}

\begin{tabular*}{\textwidth}{@{\extracolsep{\fill}}c}
\multicolumn{1}{p{\textwidth}}{\scriptsize \emph{Notes:} In four cases (Australia–Britain–Canada–Israel–US combined, Colombia, Norway, and the UK), only one study is available for the country, resulting in weighted and unweighted means being identical.  Weighted = coefficients are weighted by the inverse of the number of estimates reported per study. The column $\sigma$ depicts implied elasticity computed as $\sigma = -1/mean$ using the coefficient values prior to rounding. N is the number of observations per sample. No winsorization needed.}
\end{tabular*}
\end{center}
\vspace{-0.4cm}
\end{table}
\end{singlespace}

\newpage

\section{Heterogeneity}
\label{sec:apphet}
\setcounter{table}{0}
\setcounter{figure}{0}

\begin{spacing}{1.0}
\begin{longtable}{@{\extracolsep{\fill}}l>{\raggedright\arraybackslash}p{6cm}ccc}
\caption{Description and summary statistics of regression variables} \label{tab:descr} \\
\toprule
Variable                 & Description                                                                                                                                                                                                                                   & Mean   & SD & WM     \\
\midrule
\endfirsthead

\multicolumn{5}{c}%
{{\bfseries \tablename\ \thetable{} -- continued from previous page}} \\
\toprule
Variable                 & Description                                                                                                                                                                                                                                   & Mean   & SD & WM     \\
\midrule
\endhead

\midrule \multicolumn{5}{r}{{Continued on next page}} \\ 
\endfoot

\bottomrule
\endlastfoot

Elasticity                                        & Estimate of the negative of the inverse elasticity of substitution between the immigrant and native labor (response variable).                                                                                      & -0.08 & 0.11      & -0.08 \\
\sym{*} Standard error (SE)      & Standard error of the estimated inverse elasticity. The variable is important for gauging publication bias.                                                                                                         & 0.03  & 0.04      & 0.04  \\
\textbf{Data characteristics}                              &                                                                                                                                                                                                                     &       &           &       \\
\sym{*} Number of cells          & The logarithm of the total number of the multidimensional partitions of the labor market usually based on education, experience, region, gender, and time.                                                                                                            & 2.28  & 0.61      & 2.37  \\
\sym{*} Annual frequency data    & = 1 if annual frequency of the data is used in the estimation, reference category; = 0 if decennial or 5 year span.                                                                                                 & 0.45  & 0.50      & 0.54  \\
\textbf{Structural variation}                              &                                                                                                                                                                                                                     &       &           &       \\
\sym{*} All workers              & = 1 if all workers included (including self-employed, in-school, or part-time workers), reference category; = 0 if only full-time workers included.                                                                 & 0.55  & 0.50      & 0.56  \\
\sym{*} High level of experience & = 1 if only high experienced workers included in the estimation (years of experience equal to or higher than 20).                                                                                                   & 0.04  & 0.19      & 0.02  \\
\sym{*} Low level of experience  & = 1 if only low experienced workers included in the estimation (years of experience lower than 20).                                                                                                                 & 0.07  & 0.26      & 0.03  \\
\sym{*} High level of education  & = 1 if only high educated workers included in the estimation (worker has some college degree).                                                                                                                      & 0.10  & 0.30      & 0.11  \\
\sym{*} Low level of education   & = 1 if only low educated workers included in the estimation (worker does not have any college degree, i.e. high school graduates as well as high school dropouts).                                                  & 0.16  & 0.36      & 0.12  \\
English                                           & = 1 if English is official language in the country; = 0 otherwise.                                                                                                                                                  & 0.67  & 0.47      & 0.63  \\
Bilingual                                         & = 1 if the country for which the elasticity is estimated has two or more official languages (multilingualism); = 0 otherwise.                                                                                       & 0.12  & 0.33      & 0.11  \\
\sym{*} Top 6 languages          & = 1 if one of the official languages of the country is in top 6 languages by total number of speakers (English, Mandarin Chinese, Hindi, Spanish, Standard Arabic, French), reference category; = 0 otherwise.      & 0.84  & 0.37      & 0.80  \\
\sym{*} Male                     & = 1 if only men included in the estimation.                                                                                                                                                                         & 0.41  & 0.49      & 0.43  \\
\sym{*} Female                   & = 1 if only women included in the estimation.                                                                                                                                                                        & 0.09  & 0.29      & 0.05  \\
Both                                              & = 1 if women and men (pooled) included in the estimation.                                                                                                                                                           & 0.50  & 0.50      & 0.52  \\
Farmers                                           & = 1 if only farmers included in the estimation.                                                                                                                                                                     & 0.03  & 0.17      & 0.05  \\
North America                                     & = 1 if the country for which the elasticity is estimated is the United States or Canada; = 0 otherwise.                                                                                                             & 0.65  & 0.48      & 0.60  \\
\sym{*} Immigrant population               & The percentage of foreign-born population.                                                                                                                                                                          & 0.09  & 0.05      & 0.10  \\
\textbf{Estimation characteristics}              &                                                                                                                                                                                                                     &       &           &       \\
\sym{*} DV: Log of mean wages        & = 1 if Log of mean wages used as dependent variable, reference category; = 0 if Mean of log wages.                                                                                                                          & 0.79  & 0.41      & 0.60  \\
\sym{*} Annual wage              & = 1 if the study uses annual wages to examine the elasticity of substitution.                                                                                                                                       & 0.09  & 0.28      & 0.09  \\
\sym{*} Monthly wage             & = 1 if the study uses monthly wages to examine the elasticity of substitution.                                                                                                                                      & 0.10  & 0.30      & 0.08  \\
Weekly wage              & = 1 if the study uses weekly wages to examine the elasticity of substitution.                                                                                                                                       & 0.45  & 0.50      & 0.38  \\
\sym{*} Daily wage               & = 1 if the study uses daily wages to examine the elasticity of substitution.                                                                                                                                        & 0.05  & 0.21      & 0.10  \\
\sym{*} Hourly wage              & = 1 if the study uses hourly wages to examine the elasticity of substitution.                                                                                                                                       & 0.32  & 0.47      & 0.35  \\
\sym{*} National                 & = 1 if national approach, reference category; = 0 if regional approach.                                                                                                                                             & 0.78  & 0.42      & 0.73  \\
\sym{*} OLS                      & =1 if the ordinary least squares method, its variations (LS, DOLS, WLS, GLS) or LASSO estimates are used for estimation, reference category; = 0 if instrumental variables are used, including 2SLS, 3SLS, and GMM. & 0.73  & 0.45      & 0.74  \\
\sym{*} Time fixed effects & = 1 if time fixed effects are included in the model.                                                                                                                                                                & 0.72  & 0.45      & 0.74  \\
\sym{*} Person fixed effects & = 1 if the model includes fixed effects for individual characteristics (age, location, gender, \ldots).                                                                                                                                  & 0.23  & 0.42      & 0.34  \\
\sym{*} Skill fixed effects & = 1 if skill fixed effects are included in the model (education, experience).                                                                                                                                       & 0.88  & 0.33      & 0.84  \\
\textbf{Publication characteristics}                       &                                                                                                                                                                                                                     &       &           &       \\
\sym{*} Impact factor            & The discounted recursive RePEc impact factor of the outlet (March 3, 2025).                                                                                                                                         & 0.59  & 0.91      & 0.81  \\
\sym{*} Citations                & The logarithm of the number of per-year citations of the study in Google Scholar (March 3, 2025).                                                                                                                   & 0.80  & 0.80      & 0.74  \\
\sym{*} Published                & =1 if a study is published in a peer-reviewed journal, reference category; = 0 if a study is in a form of working paper or PhD Thesis.                                                                              & 0.72  & 0.45      & 0.75  \\
Unpublished old                                   & = 1 if a working paper has remained unpublished for more than 10 years.                                                                                                                                                                     & 0.22  & 0.42      & 0.15  \\
\sym{*} Top 5 journals           & = 1 if a study published in the top 5 economic journals.                                                                                                                                                            & 0.06  & 0.24      & 0.13  \\
Top 20 journals                                   & = 1 if a study published in the top 20 economic journals.                                                                                                                                                           & 0.24  & 0.43      & 0.33  \\
\sym{*} Publication year   & The logarithm of the study's publication year. & 3.30  & 0.00      & 3.30 \\
\midrule
Observations: 1,087      &                                                                                                                                                                                                                                               &        &           &        \\
Studies: 40              &                                                                                                                                                                                                                                               &        &           &        \\
\end{longtable}
\end{spacing}
Note that the heterogeneity analysis comprises 1,087 observations (from 40 studies). The number of observations differs slightly from the publication bias analysis. We excluded 4 observations (1 study) because they were estimated using a multi-country sample, which precluded the assignment of country-specific moderator variables, such as the immigration share. The summary statistics of the whole dataset are in \autoref{tab:tab1}. SD = standard deviation, WM = mean weighted by the inverse of the number of estimates reported per study. \sym{*} indicates that the variable was used for heterogeneity analysis.

\newgeometry{left=1.5cm,right=1.5cm,top=2cm,bottom=2cm}
\begin{landscape}
\thispagestyle{empty}
\begin{table}[htbp]
\centering
\begin{threeparttable}
\caption{Robustness of BMA results across different random-model priors}
\label{tab:bma_robust_random}
\begin{tabular}{lccccccccccccc}
\toprule
\textbf{Random model prior} & \multicolumn{3}{c}{\textbf{UIP prior}} & \multicolumn{3}{c}{\textbf{BRIC prior}} & \multicolumn{3}{c}{\textbf{Hyper-g prior}} & \multicolumn{3}{c}{\textbf{EBL prior}} \\
\cmidrule(lr){2-4} \cmidrule(lr){5-7} \cmidrule(lr){8-10} \cmidrule(lr){11-13}
 & PIP & Post Mean & Post SD & PIP & Post Mean & Post SD & PIP & Post Mean & Post SD & PIP & Post Mean & Post SD \\
\midrule
Constant & 1.00 & 1.05 & NA & 1.00 & 1.04 & NA & 1.00 & 6.28 & NA & 1.00 & 6.30 & NA \\
SE & 1.00 & -0.52 & 0.08 & 1.00 & -0.52 & 0.08 & 1.00 & -0.52 & 0.08 & 1.00 & -0.52 & 0.08 \\
Number of cells & 1.00 & 0.06 & 0.01 & 1.00 & 0.06 & 0.01 & 1.00 & 0.06 & 0.01 & 1.00 & 0.06 & 0.01 \\
Annual frequency data & 1.00 & -0.08 & 0.01 & 1.00 & -0.08 & 0.01 & 1.00 & -0.08 & 0.01 & 1.00 & -0.08 & 0.01 \\
All workers & 1.00 & -0.05 & 0.01 & 1.00 & -0.05 & 0.01 & 1.00 & -0.05 & 0.01 & 1.00 & -0.05 & 0.01 \\
High level of experience & 0.05 & 0.00 & 0.00 & 0.05 & 0.00 & 0.00 & 0.47 & 0.00 & 0.01 & 0.47 & 0.00 & 0.01 \\
Low level of experience & 1.00 & -0.05 & 0.01 & 1.00 & -0.05 & 0.01 & 1.00 & -0.04 & 0.01 & 1.00 & -0.04 & 0.01 \\
High level of education & 0.06 & 0.00 & 0.00 & 0.06 & 0.00 & 0.00 & 0.49 & 0.00 & 0.01 & 0.49 & 0.00 & 0.01 \\
Low level of education & 0.05 & 0.00 & 0.00 & 0.05 & 0.00 & 0.00 & 0.47 & 0.00 & 0.01 & 0.47 & 0.00 & 0.01 \\
Top 6 languages & 0.59 & 0.02 & 0.02 & 0.59 & 0.02 & 0.02 & 0.92 & 0.03 & 0.02 & 0.92 & 0.03 & 0.02 \\
Male & 1.00 & 0.04 & 0.01 & 1.00 & 0.04 & 0.01 & 1.00 & 0.04 & 0.01 & 1.00 & 0.04 & 0.01 \\
Female & 0.47 & 0.01 & 0.02 & 0.47 & 0.01 & 0.02 & 0.88 & 0.02 & 0.01 & 0.88 & 0.02 & 0.01 \\
Immigrant population & 0.98 & 0.33 & 0.10 & 0.98 & 0.33 & 0.10 & 0.99 & 0.27 & 0.10 & 0.99 & 0.27 & 0.10 \\
DV: Log of mean wages  & 0.69 & -0.02 & 0.01 & 0.69 & -0.02 & 0.01 & 0.88 & -0.02 & 0.01 & 0.88 & -0.02 & 0.01 \\
Annual wage & 0.25 & 0.01 & 0.01 & 0.25 & 0.01 & 0.01 & 0.67 & 0.01 & 0.01 & 0.67 & 0.01 & 0.01 \\
Monthly wage & 0.06 & 0.00 & 0.00 & 0.06 & 0.00 & 0.00 & 0.48 & 0.00 & 0.01 & 0.48 & 0.00 & 0.01 \\
Daily wage & 0.06 & 0.00 & 0.01 & 0.06 & 0.00 & 0.01 & 0.51 & 0.01 & 0.02 & 0.51 & 0.01 & 0.02 \\
Hourly wage & 1.00 & -0.06 & 0.01 & 1.00 & -0.06 & 0.01 & 1.00 & -0.06 & 0.01 & 1.00 & -0.06 & 0.01 \\
National & 1.00 & -0.06 & 0.01 & 1.00 & -0.06 & 0.01 & 1.00 & -0.05 & 0.01 & 1.00 & -0.05 & 0.01 \\
OLS & 0.16 & 0.00 & 0.01 & 0.16 & 0.00 & 0.01 & 0.74 & -0.01 & 0.01 & 0.74 & -0.01 & 0.01 \\
Time fixed effects & 0.53 & -0.01 & 0.01 & 0.53 & -0.01 & 0.01 & 0.81 & -0.01 & 0.01 & 0.81 & -0.01 & 0.01 \\
Person fixed effects & 1.00 & -0.08 & 0.01 & 1.00 & -0.08 & 0.01 & 1.00 & -0.08 & 0.01 & 1.00 & -0.08 & 0.01 \\
Skill fixed effects & 1.00 & 0.07 & 0.01 & 1.00 & 0.07 & 0.01 & 1.00 & 0.07 & 0.01 & 1.00 & 0.07 & 0.01 \\
Impact factor & 0.06 & 0.00 & 0.00 & 0.06 & 0.00 & 0.00 & 0.53 & 0.00 & 0.01 & 0.53 & 0.00 & 0.01 \\
Citations & 1.00 & -0.02 & 0.01 & 1.00 & -0.02 & 0.01 & 1.00 & -0.02 & 0.01 & 1.00 & -0.02 & 0.01 \\
Published & 1.00 & 0.05 & 0.01 & 1.00 & 0.05 & 0.01 & 1.00 & 0.06 & 0.01 & 1.00 & 0.06 & 0.01 \\
Top 5 journals & 0.08 & 0.00 & 0.01 & 0.08 & 0.00 & 0.01 & 0.63 & 0.01 & 0.02 & 0.63 & 0.01 & 0.02 \\
Publication year & 0.09 & -0.37 & 1.61 & 0.09 & -0.37 & 1.61 & 0.56 & -1.96 & 3.23 & 0.56 & -1.96 & 3.23 \\
\midrule
Studies & 40 & & & & & & & & & & & \\
Observations & 1,087 &  & & & & & & & & & & \\
\bottomrule
\end{tabular}

\begin{tablenotes}
\footnotesize
\item \emph{Notes:} PIP = Posterior Inclusion Probability. SD = Standard Deviation. Results reflect Bayesian Model Averaging using random-model priors under four alternative $g$-prior formulations: UIP, BRIC, hyper-$g$, and EBL. For definitions of included covariates, see \autoref{tab:descr}. \textit{Number of cells}, \textit{citations}, and \textit{publication year} enter in logarithms.
\end{tablenotes}

\end{threeparttable}
\end{table}
\end{landscape}
\restoregeometry
\clearpage

\clearpage
\newgeometry{left=1.5cm,right=1.5cm,top=2cm,bottom=2cm}
\begin{landscape}
\thispagestyle{empty}
\begin{table}[htbp]
\centering
\begin{threeparttable}
\caption{Robustness of BMA results across different uniform-model priors}
\label{tab:bma_robust_uniform}
\begin{tabular}{lccccccccccccc}
\toprule
\textbf{Uniform model prior} & \multicolumn{3}{c}{\textbf{UIP prior}} & \multicolumn{3}{c}{\textbf{BRIC prior}} & \multicolumn{3}{c}{\textbf{Hyper-g prior}} & \multicolumn{3}{c}{\textbf{EBL prior}} \\
\cmidrule(lr){2-4} \cmidrule(lr){5-7} \cmidrule(lr){8-10} \cmidrule(lr){11-13}
 & PIP & Post Mean & Post SD & PIP & Post Mean & Post SD & PIP & Post Mean & Post SD & PIP & Post Mean & Post SD \\
\midrule
Constant & 1.00 & 0.64 & NA & 1.00 & 0.64 & NA & 1.00 & 2.64 & NA & 1.00 & 2.65 & NA \\
SE & 1.00 & -0.52 & 0.08 & 1.00 & -0.52 & 0.08 & 1.00 & -0.51 & 0.08 & 1.00 & -0.51 & 0.08 \\
Number of cells & 1.00 & 0.06 & 0.01 & 1.00 & 0.06 & 0.01 & 1.00 & 0.06 & 0.01 & 1.00 & 0.06 & 0.01 \\
Annual frequency data & 1.00 & -0.08 & 0.01 & 1.00 & -0.08 & 0.01 & 1.00 & -0.08 & 0.01 & 1.00 & -0.08 & 0.01 \\
All workers & 1.00 & -0.05 & 0.01 & 1.00 & -0.05 & 0.01 & 1.00 & -0.05 & 0.01 & 1.00 & -0.05 & 0.01 \\
High level of experience & 0.03 & 0.00 & 0.00 & 0.03 & 0.00 & 0.00 & 0.14 & 0.00 & 0.01 & 0.14 & 0.00 & 0.01 \\
Low level of experience & 1.00 & -0.05 & 0.01 & 1.00 & -0.05 & 0.01 & 1.00 & -0.05 & 0.01 & 1.00 & -0.05 & 0.01 \\
High level of education & 0.04 & 0.00 & 0.00 & 0.04 & 0.00 & 0.00 & 0.15 & 0.00 & 0.00 & 0.15 & 0.00 & 0.00 \\
Low level of education & 0.03 & 0.00 & 0.00 & 0.03 & 0.00 & 0.00 & 0.14 & 0.00 & 0.00 & 0.14 & 0.00 & 0.00 \\
Top 6 languages & 0.54 & 0.02 & 0.02 & 0.54 & 0.02 & 0.02 & 0.76 & 0.02 & 0.02 & 0.76 & 0.02 & 0.02 \\
Male & 1.00 & 0.04 & 0.01 & 1.00 & 0.04 & 0.01 & 1.00 & 0.04 & 0.01 & 1.00 & 0.04 & 0.01 \\
Female & 0.39 & 0.01 & 0.01 & 0.39 & 0.01 & 0.01 & 0.68 & 0.02 & 0.01 & 0.68 & 0.02 & 0.01 \\
Immigrant population & 0.98 & 0.34 & 0.10 & 0.98 & 0.34 & 0.10 & 0.98 & 0.31 & 0.10 & 0.98 & 0.31 & 0.10 \\
DV: Log of mean wages  & 0.64 & -0.01 & 0.01 & 0.64 & -0.01 & 0.01 & 0.78 & -0.02 & 0.01 & 0.79 & -0.02 & 0.01 \\
Annual wage & 0.21 & 0.01 & 0.01 & 0.21 & 0.01 & 0.01 & 0.39 & 0.01 & 0.01 & 0.39 & 0.01 & 0.01 \\
Monthly wage & 0.04 & 0.00 & 0.00 & 0.04 & 0.00 & 0.00 & 0.15 & 0.00 & 0.01 & 0.15 & 0.00 & 0.01 \\
Daily wage & 0.04 & 0.00 & 0.00 & 0.04 & 0.00 & 0.00 & 0.17 & 0.00 & 0.01 & 0.17 & 0.00 & 0.01 \\
Hourly wage & 1.00 & -0.06 & 0.01 & 1.00 & -0.06 & 0.01 & 1.00 & -0.06 & 0.01 & 1.00 & -0.06 & 0.01 \\
National & 1.00 & -0.06 & 0.01 & 1.00 & -0.06 & 0.01 & 1.00 & -0.06 & 0.01 & 1.00 & -0.06 & 0.01 \\
OLS & 0.11 & 0.00 & 0.00 & 0.11 & 0.00 & 0.00 & 0.38 & 0.00 & 0.01 & 0.38 & 0.00 & 0.01 \\
Time fixed effects & 0.47 & -0.01 & 0.01 & 0.47 & -0.01 & 0.01 & 0.66 & -0.01 & 0.01 & 0.67 & -0.01 & 0.01 \\
Person fixed effects & 1.00 & -0.08 & 0.01 & 1.00 & -0.08 & 0.01 & 1.00 & -0.08 & 0.01 & 1.00 & -0.08 & 0.01 \\
Skill fixed effects & 1.00 & 0.07 & 0.01 & 1.00 & 0.07 & 0.01 & 1.00 & 0.07 & 0.01 & 1.00 & 0.07 & 0.01 \\
Impact factor & 0.04 & 0.00 & 0.00 & 0.04 & 0.00 & 0.00 & 0.17 & 0.00 & 0.00 & 0.17 & 0.00 & 0.00 \\
Citations & 1.00 & -0.02 & 0.01 & 1.00 & -0.02 & 0.01 & 1.00 & -0.02 & 0.01 & 1.00 & -0.02 & 0.01 \\
Published & 1.00 & 0.05 & 0.01 & 1.00 & 0.05 & 0.01 & 1.00 & 0.05 & 0.01 & 1.00 & 0.05 & 0.01 \\
Top 5 journals & 0.05 & 0.00 & 0.01 & 0.05 & 0.00 & 0.01 & 0.24 & 0.00 & 0.01 & 0.24 & 0.00 & 0.01 \\
Publication year & 0.06 & -0.25 & 1.32 & 0.06 & -0.25 & 1.32 & 0.23 & -0.85 & 2.36 & 0.23 & -0.86 & 2.36 \\
\midrule
Studies &  40 && & & & & & & & & & \\
Observations & 1,087 & & & & & & & & & & & \\
\bottomrule
\end{tabular}
\begin{tablenotes}

\footnotesize
\item \emph{Notes:} PIP = Posterior Inclusion Probability. SD = Standard Deviation. Results reflect Bayesian Model Averaging using uniform-model priors under four alternative $g$-prior formulations: UIP, BRIC, hyper-$g$, and EBL. For definitions of included covariates, see \autoref{tab:descr}. \textit{Number of cells}, \textit{citations}, and \textit{publication year} enter in logarithms.
\end{tablenotes}

\end{threeparttable}
\end{table}
\end{landscape}
\restoregeometry
\clearpage

\begin{figure}[h!]
    \centering
    % First row of images
    \begin{minipage}[t]{0.48\textwidth}
        \centering
        \includegraphics[width=\textwidth]{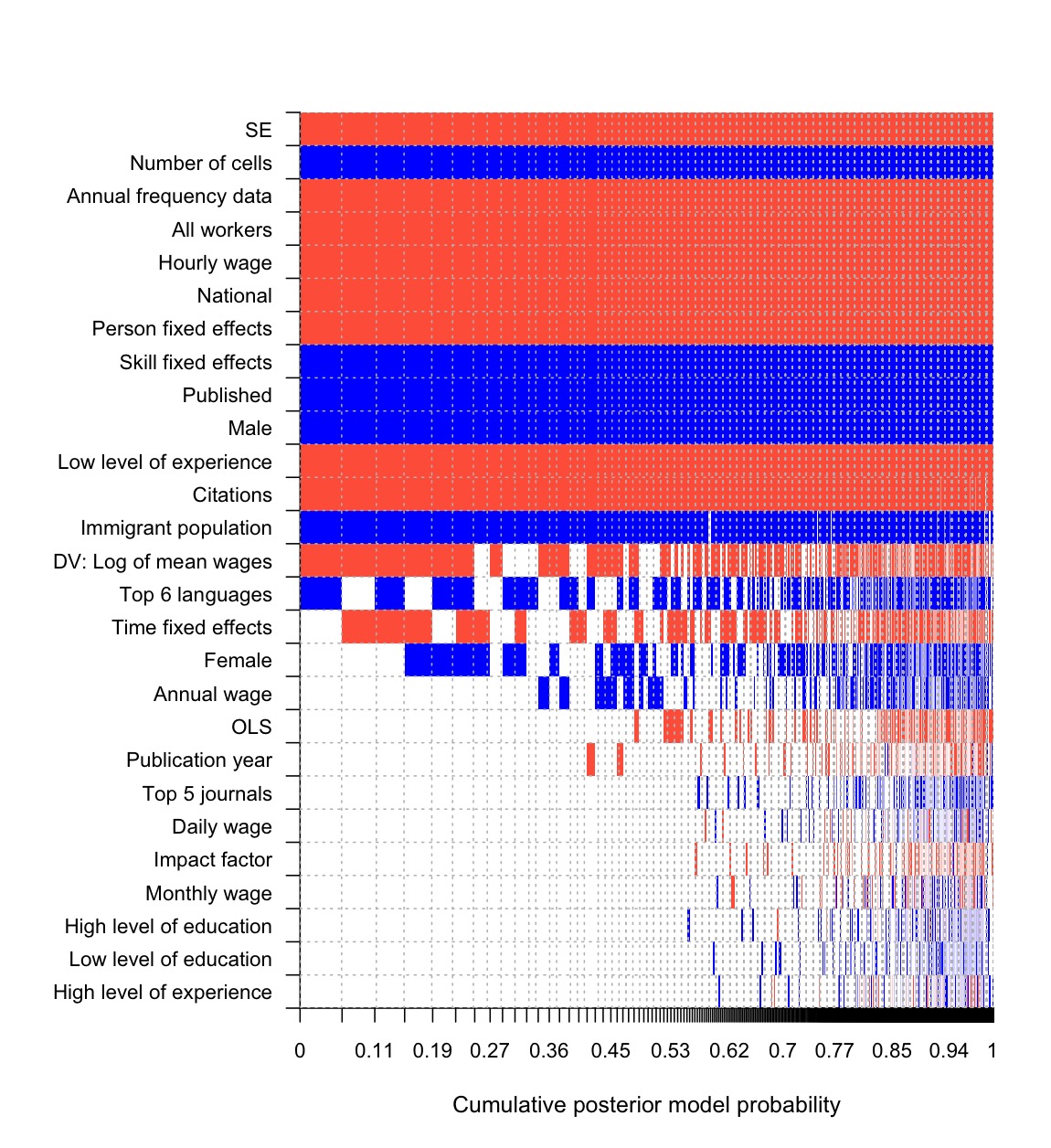}
        \caption{BMA: UIP and random priors}
        \label{fig:BMA_UIP_random}
    \end{minipage}
    \hfill
    \begin{minipage}[t]{0.48\textwidth}
        \centering
        \includegraphics[width=\textwidth]{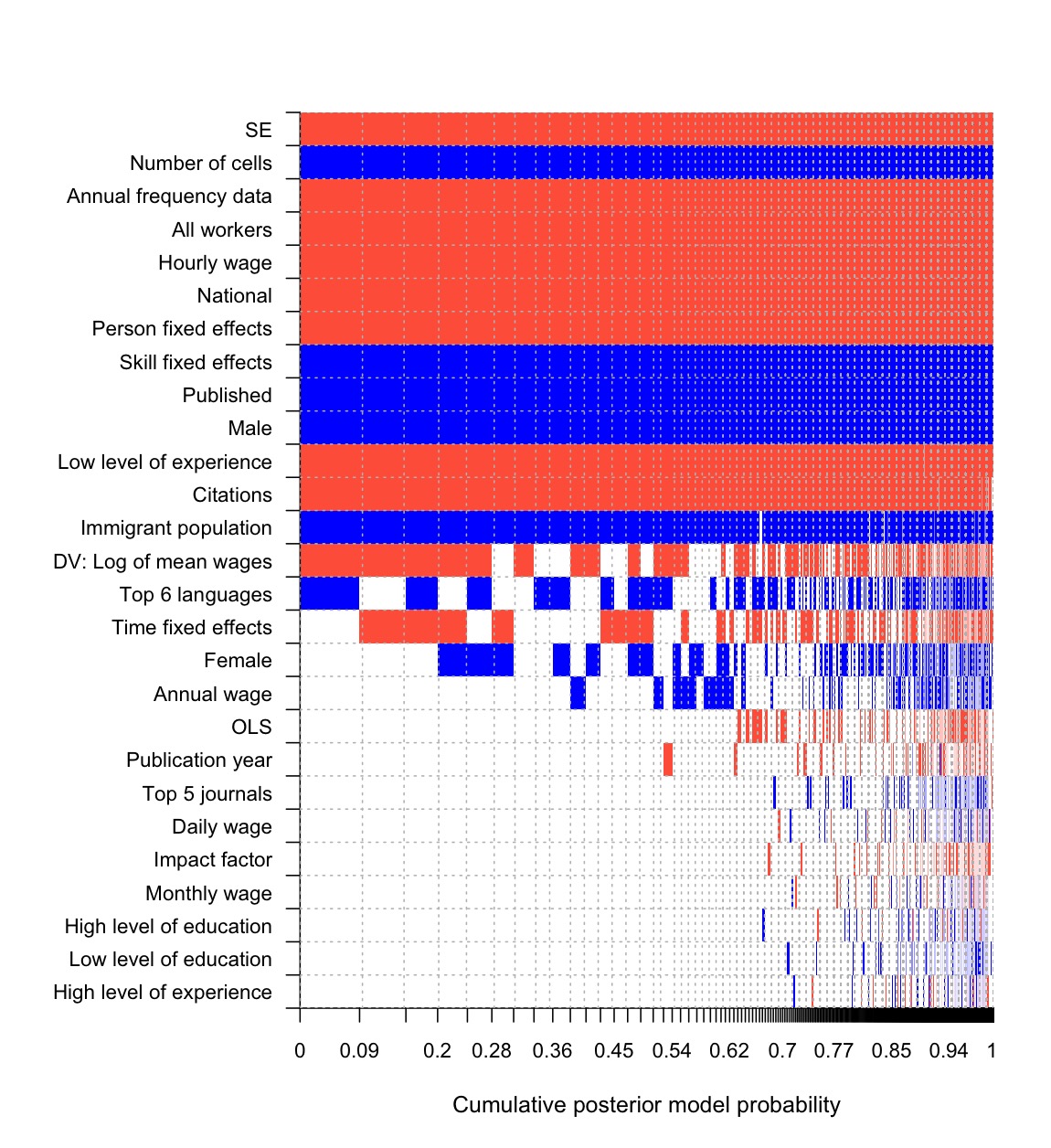}
        \caption{BMA: UIP and uniform priors}
        \label{fig:BMA_UIP_uniform}
    \end{minipage}

    % Add some vertical spacing between rows
    \vspace{0.3cm}

    % Second row of images
    \begin{minipage}[t]{0.48\textwidth}
        \centering
        \includegraphics[width=\textwidth]{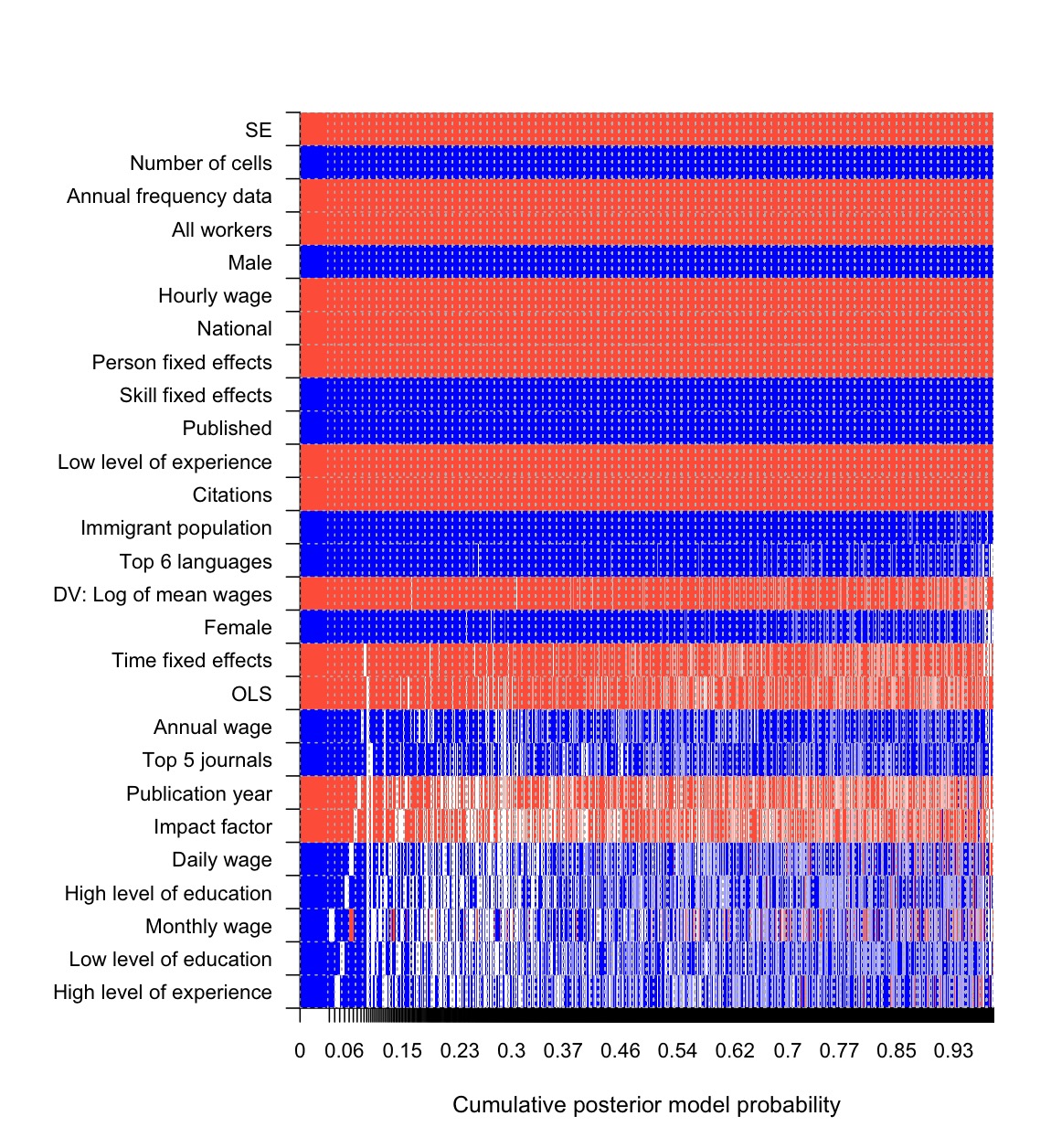}
        \caption{BMA: hyper-g and random priors}
        \label{fig:BMA_hyper_random}
    \end{minipage}
    \hfill
    \begin{minipage}[t]{0.48\textwidth}
        \centering
        \includegraphics[width=\textwidth]{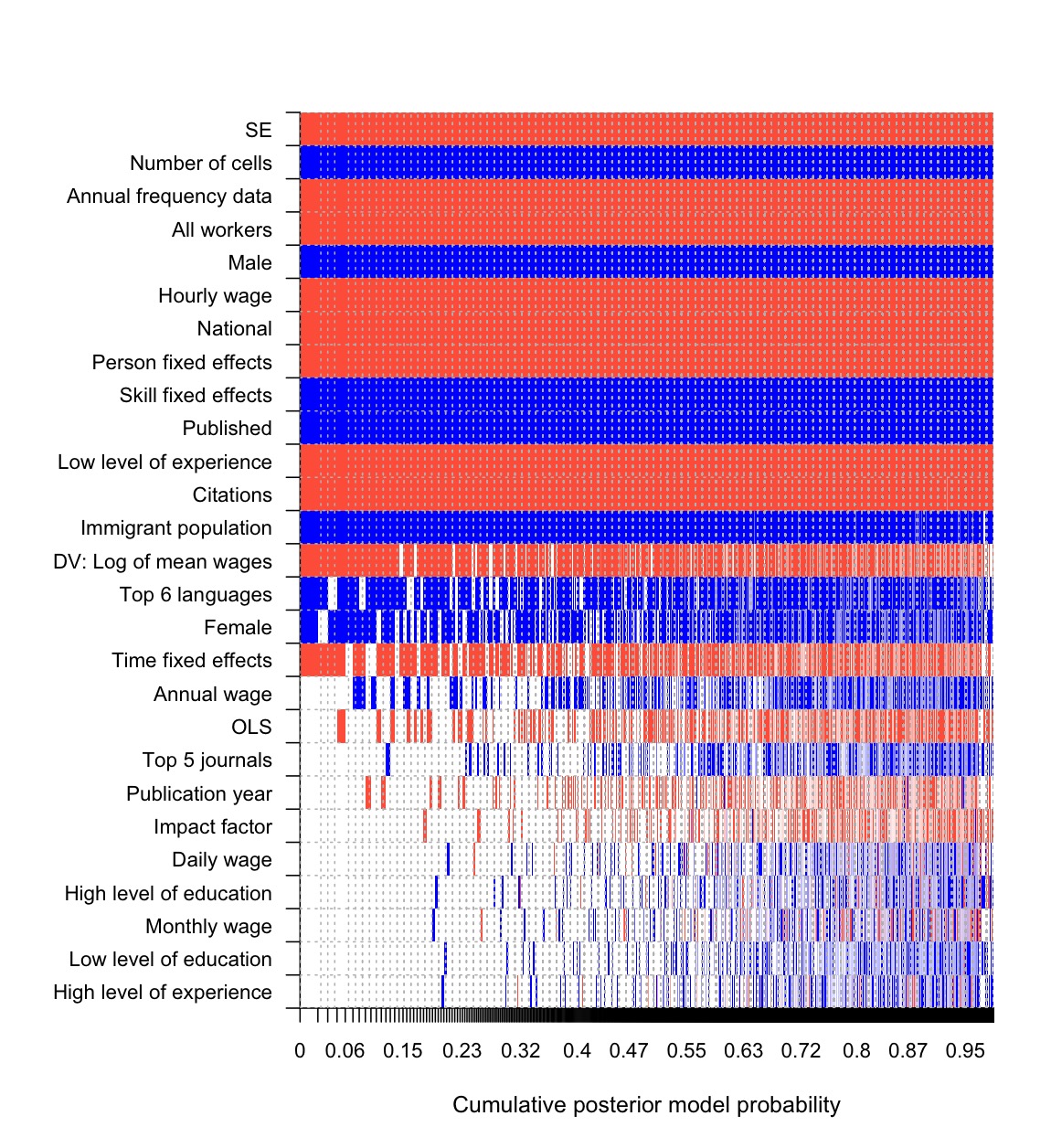}
        \caption{BMA: hyper-g and uniform priors}
        \label{fig:BMA_hyper_uniform}
    \end{minipage}
\end{figure}

\begin{figure}[h!]
    \centering
    % First row of images
    \begin{minipage}[t]{0.48\textwidth}
        \centering
        \includegraphics[width=\textwidth]{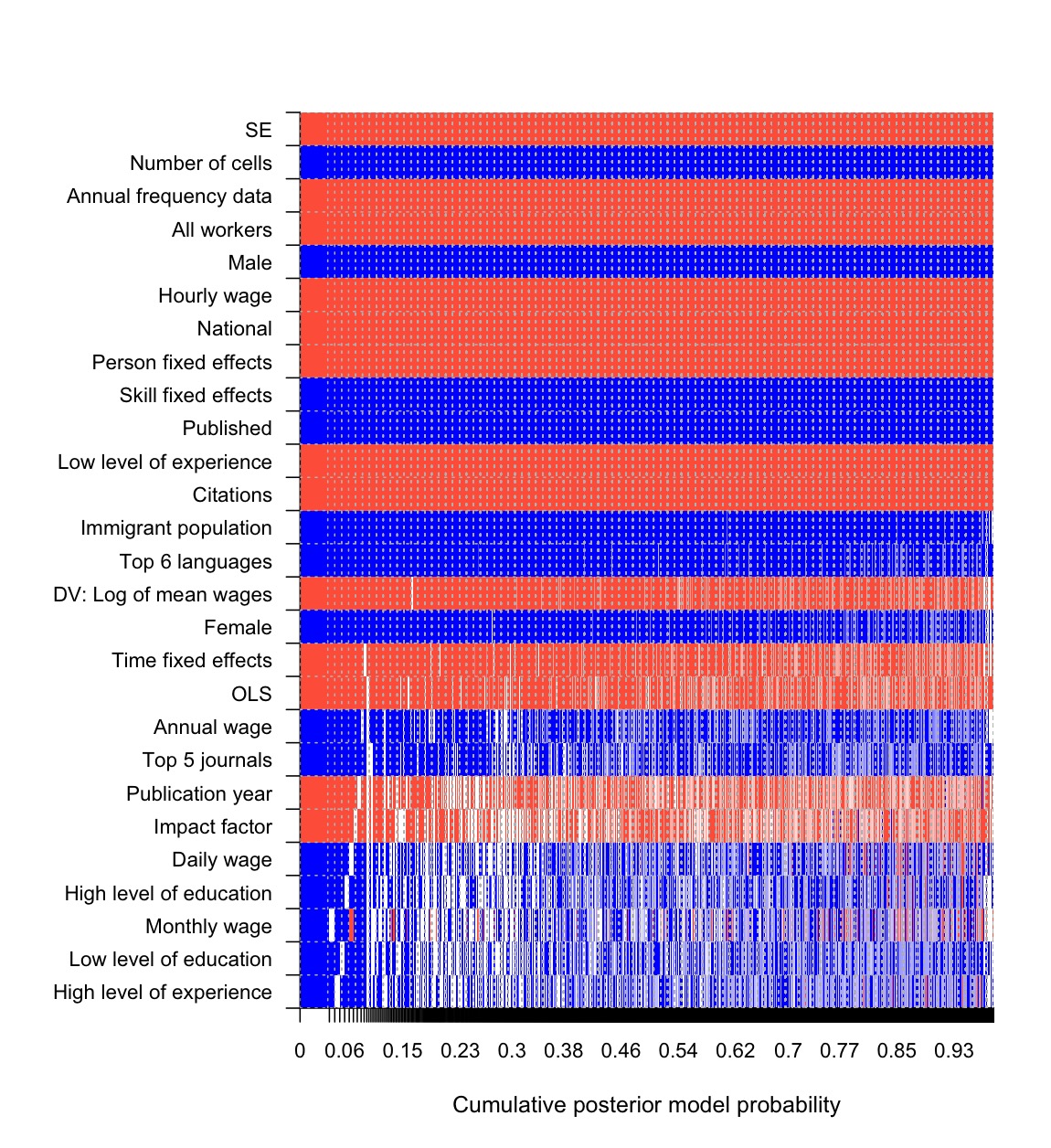}
        \caption{BMA: EBL and random priors}
        \label{fig:BMA_EBL_random}
    \end{minipage}
    \hfill
    \begin{minipage}[t]{0.48\textwidth}
        \centering
        \includegraphics[width=\textwidth]{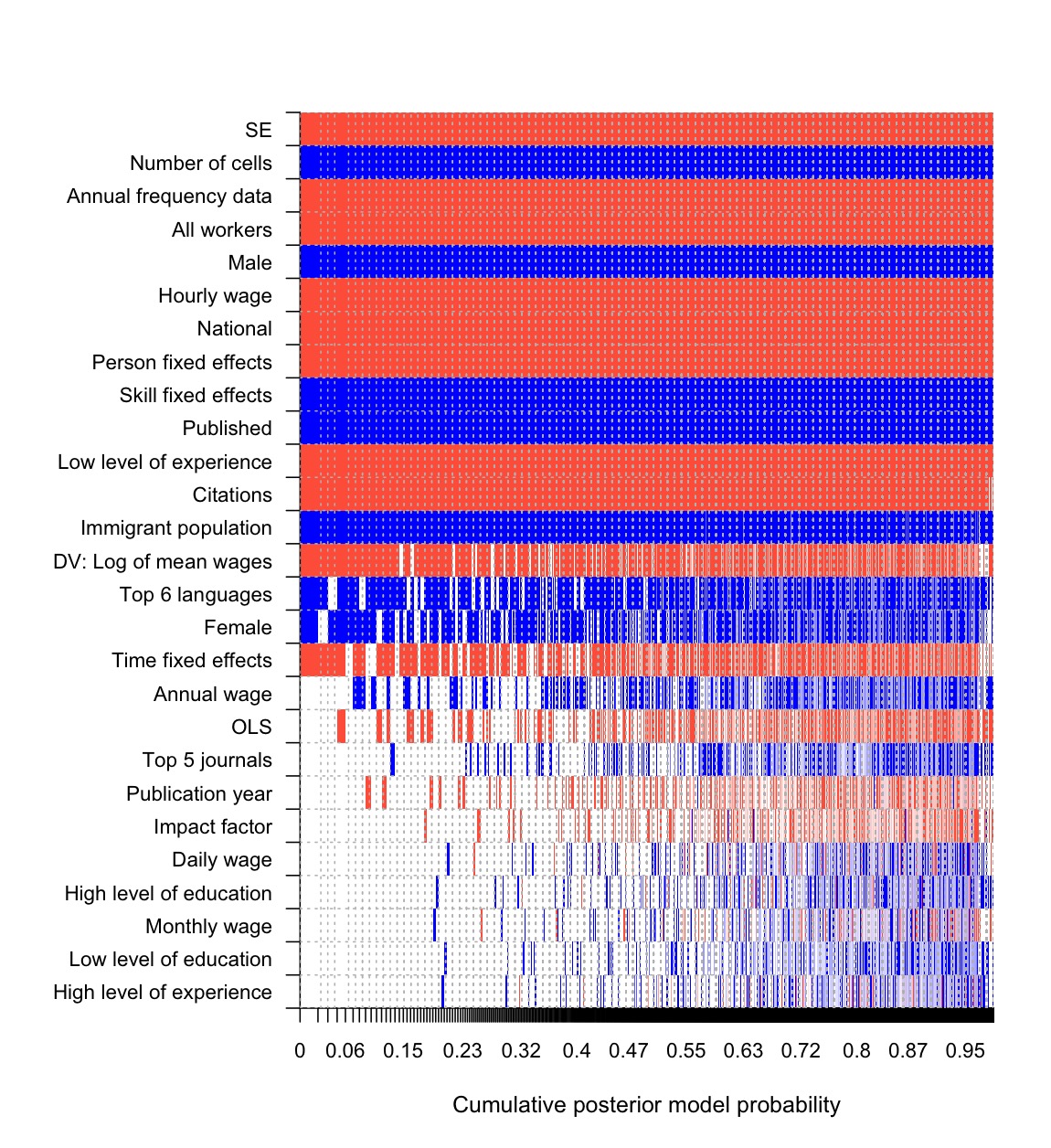}
        \caption{BMA: EBL and uniform priors}
        \label{fig:BMA_EBL_uniform}
    \end{minipage}

    % Add some vertical spacing between rows
    \vspace{0.3cm}

    % Second row of images
    \begin{minipage}[t]{0.48\textwidth}
        \centering
        \includegraphics[width=\textwidth]{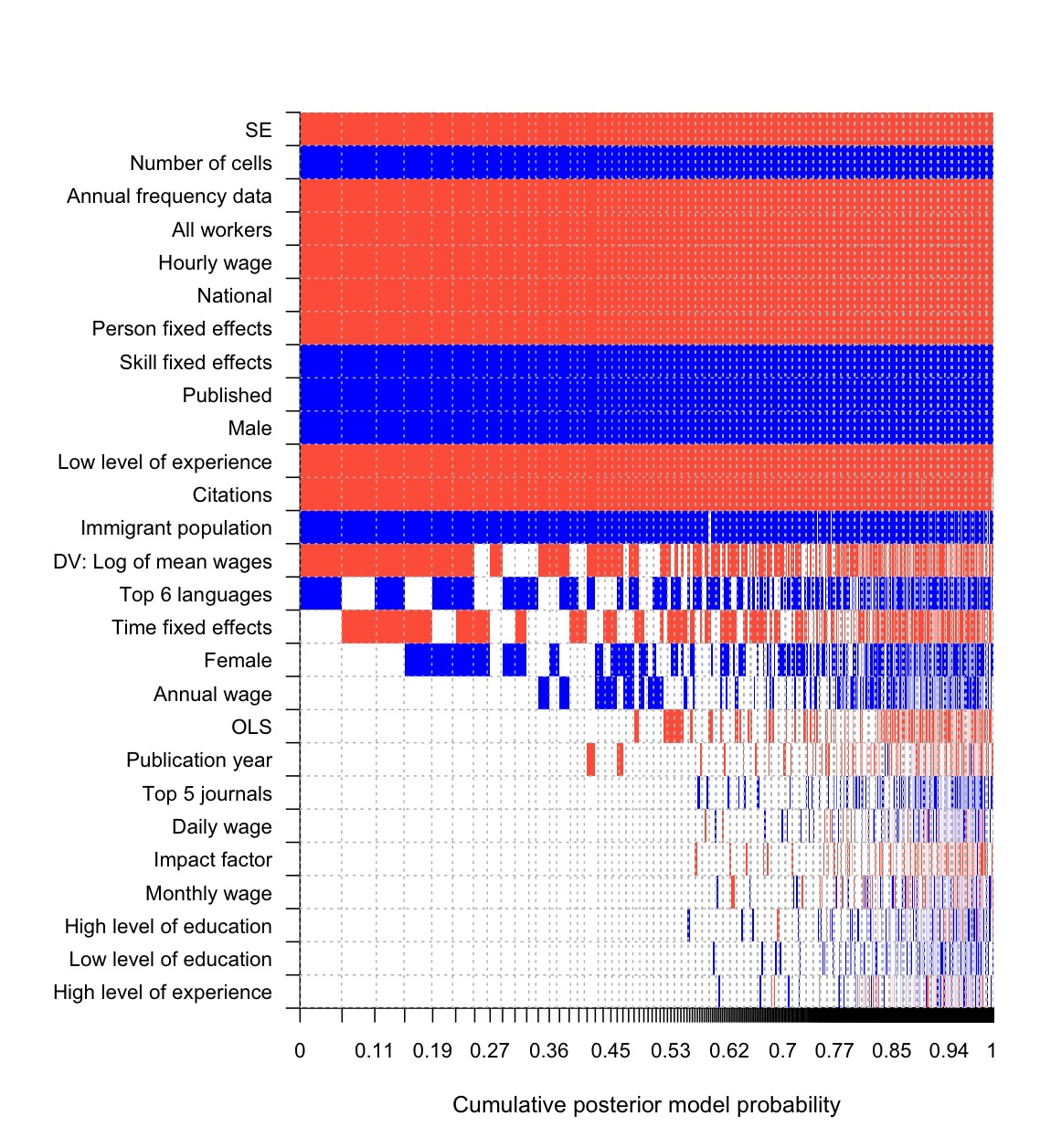}
        \caption{BMA: BRIC and random priors}
        \label{fig:BMA_BRIC_random}
    \end{minipage}
    \hfill
    \begin{minipage}[t]{0.48\textwidth}
        \centering
        \includegraphics[width=\textwidth]{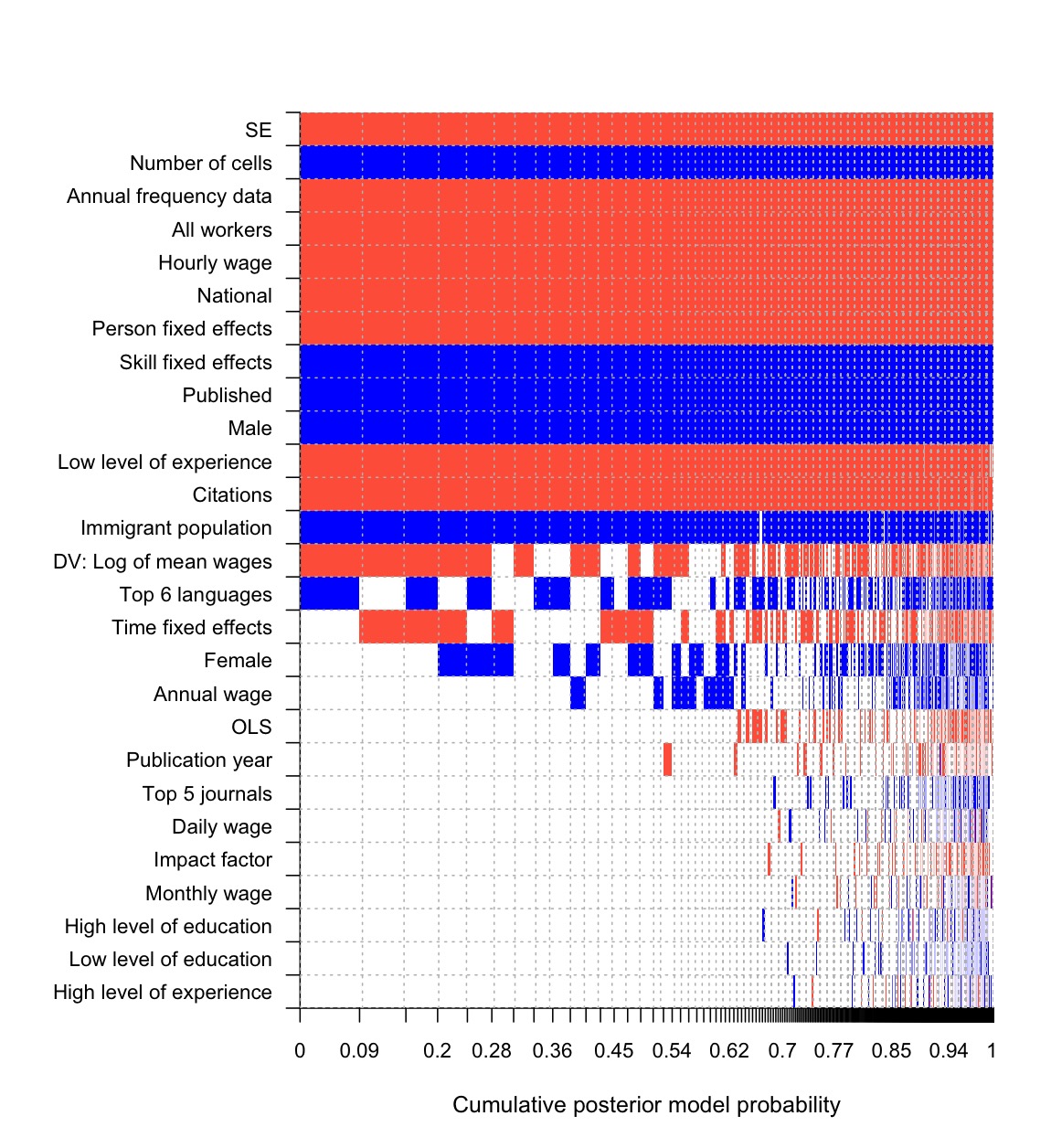}
        \caption{BMA: BRIC and uniform priors}
        \label{fig:BMA_BRIC_uniform}
    \end{minipage}
\end{figure}

\begin{figure}[h!]
    \centering
    % First row of images
    \begin{minipage}[t]{0.48\textwidth}
        \centering
        \includegraphics[width=\textwidth]{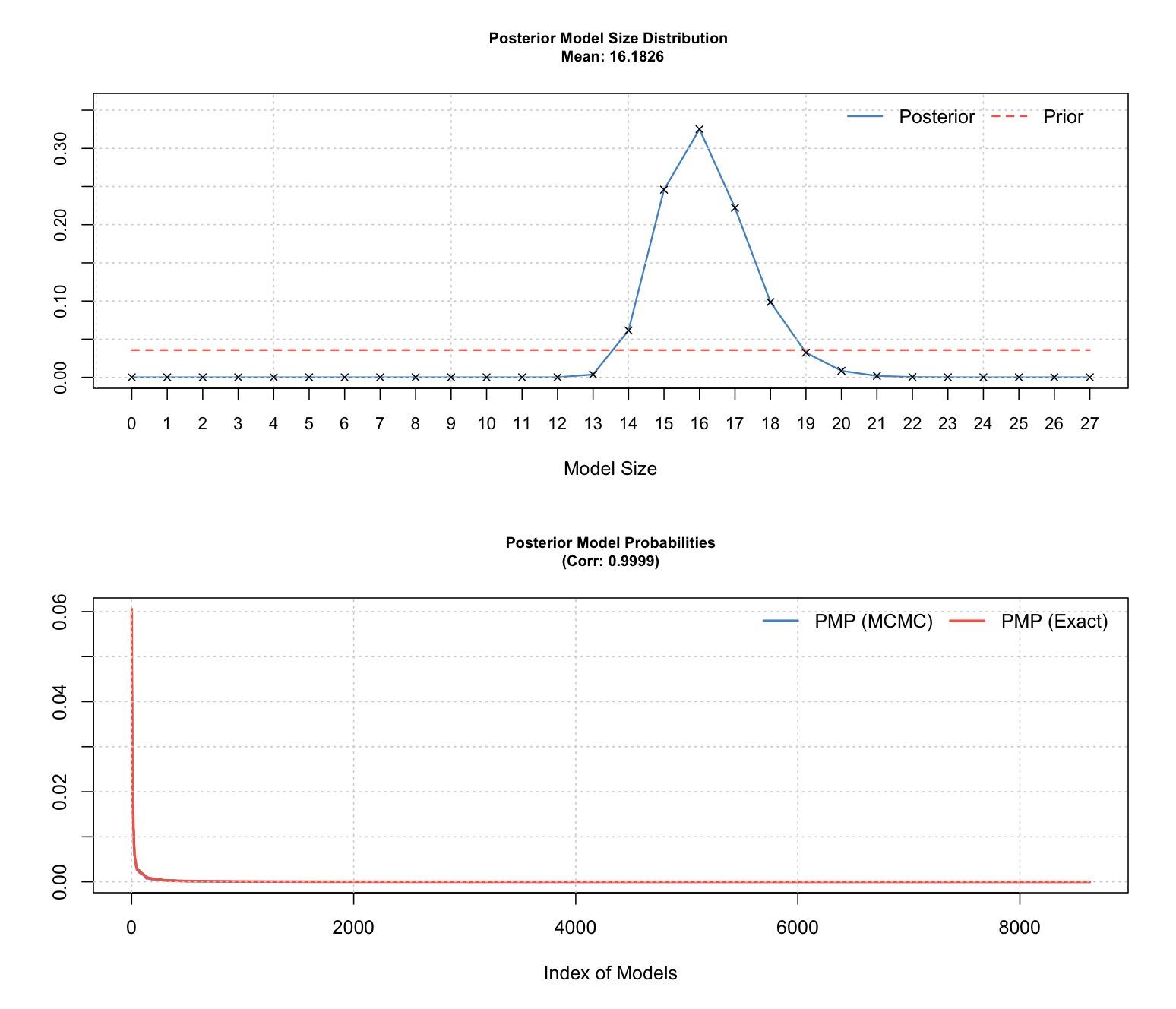}
        \caption{UIP and random priors}
        \label{fig:PMP_UIP_random}
    \end{minipage}
    \hfill
    \begin{minipage}[t]{0.48\textwidth}
        \centering
        \includegraphics[width=\textwidth]{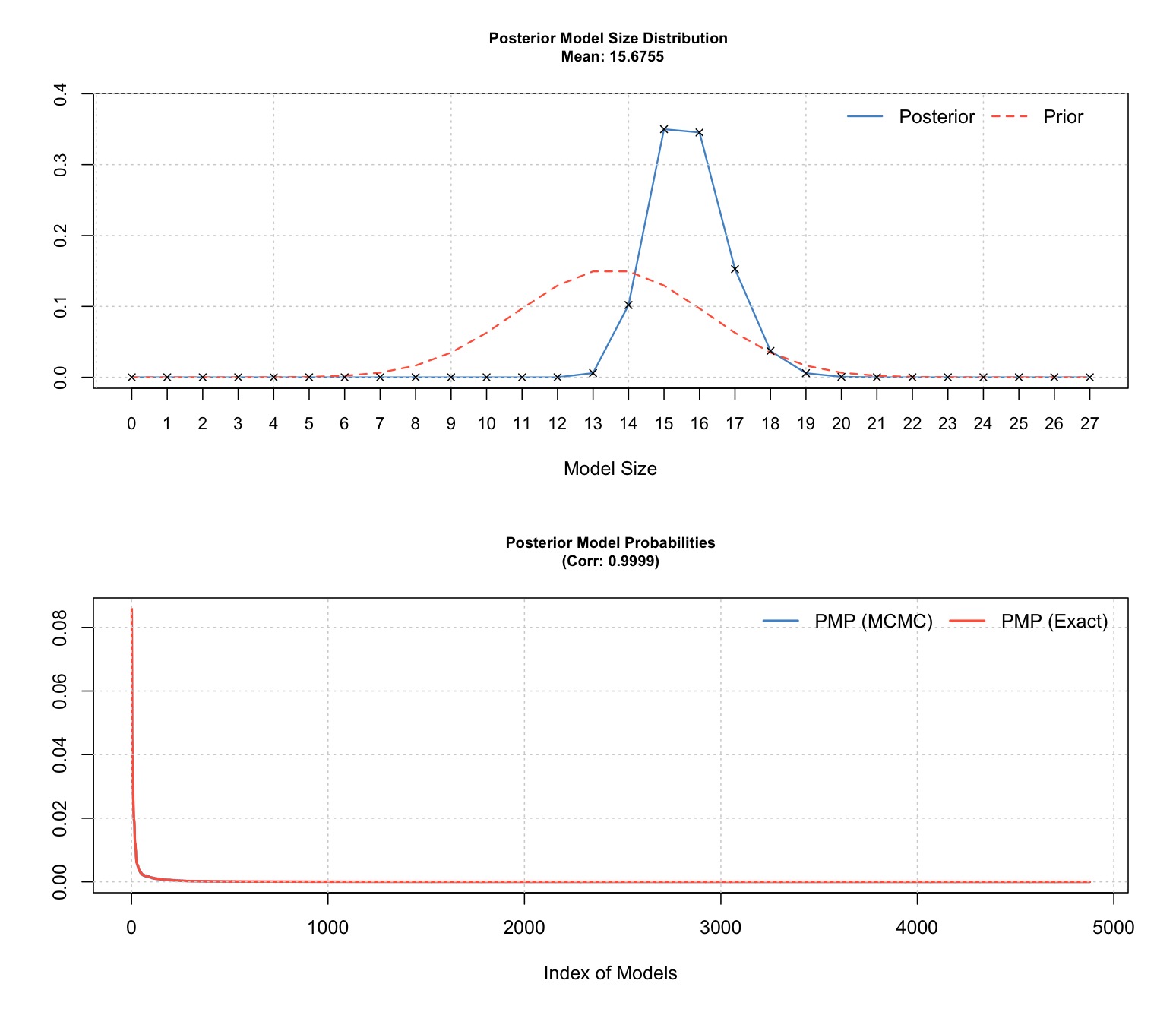}
        \caption{UIP and uniform priors}
        \label{fig:PMP_UIP_uniform}
    \end{minipage}

    % Add some vertical spacing between rows
    \vspace{0.3cm}

    % Second row of images
    \begin{minipage}[t]{0.48\textwidth}
        \centering
        \includegraphics[width=\textwidth]{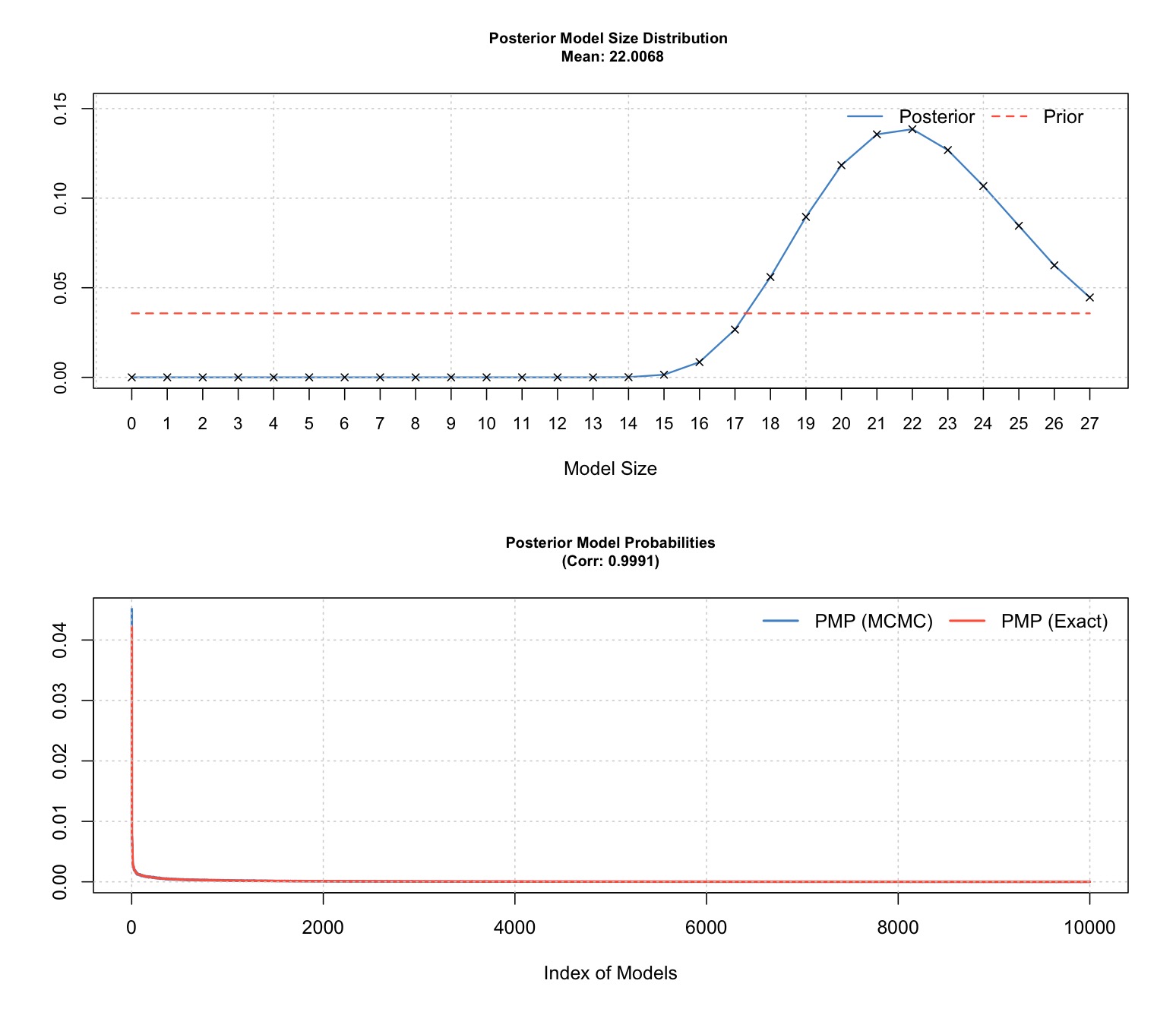}
        \caption{hyper-g and random priors}
        \label{fig:PMP_hyper_random}
    \end{minipage}
    \hfill
    \begin{minipage}[t]{0.48\textwidth}
        \centering
        \includegraphics[width=\textwidth]{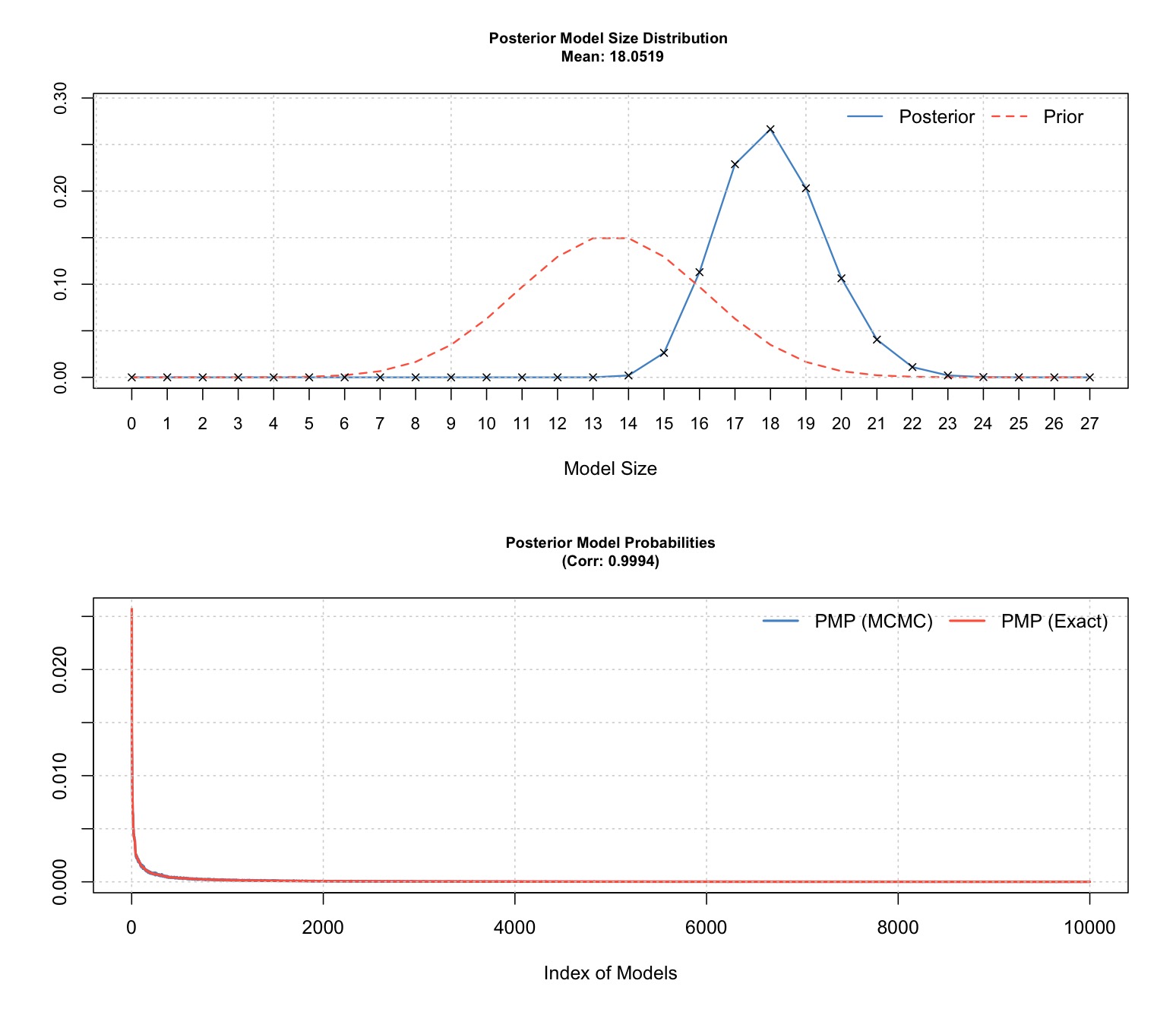}
        \caption{hyper-g and uniform priors}
        \label{fig:PMP_hyper_uniform}
    \end{minipage}
\end{figure}

\begin{figure}[h!]
    \centering
    % First row of images
    \begin{minipage}[t]{0.48\textwidth}
        \centering
        \includegraphics[width=\textwidth]{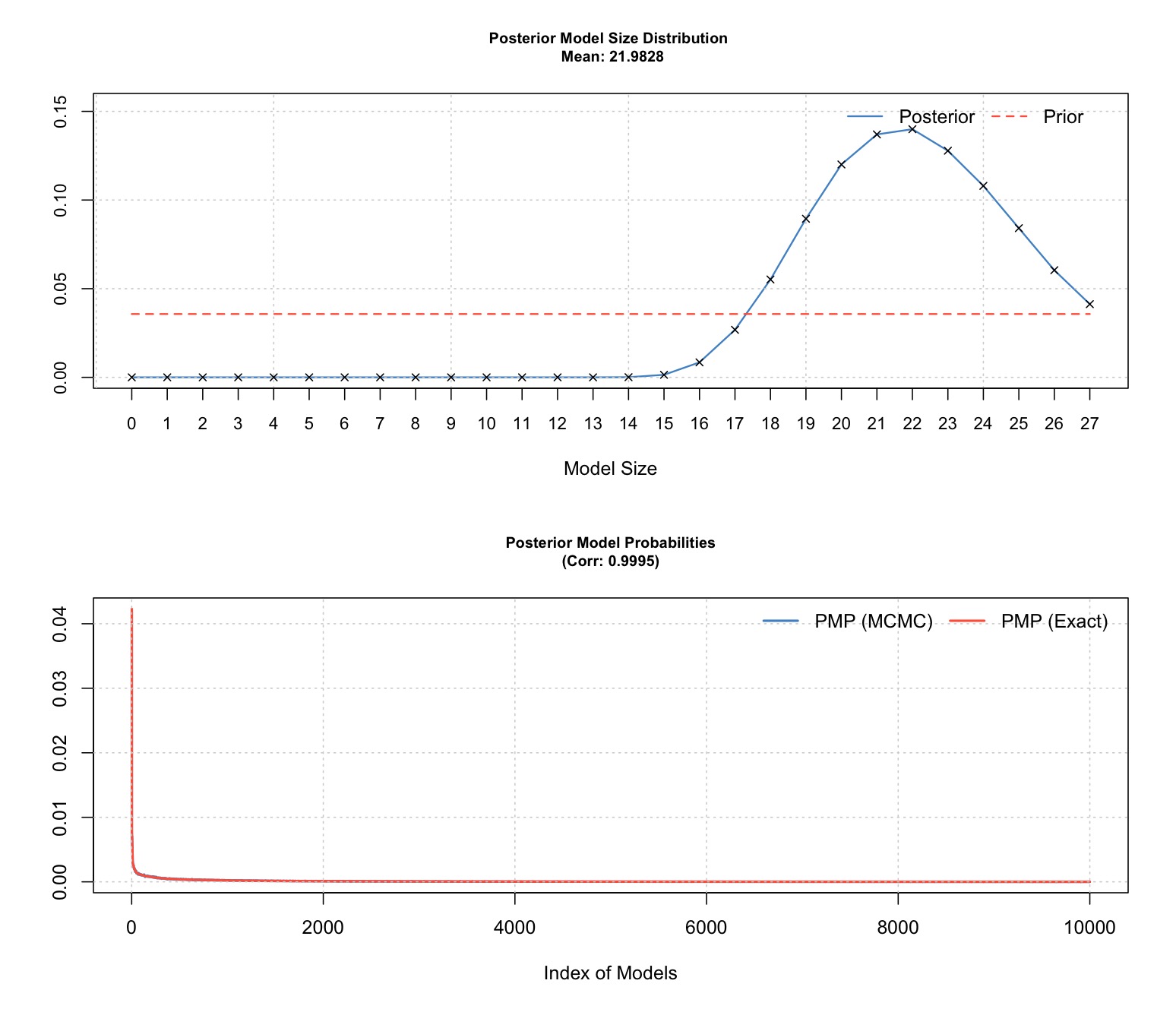}
        \caption{EBL and random priors}
        \label{fig:PMP_EBL_random}
    \end{minipage}
    \hfill
    \begin{minipage}[t]{0.48\textwidth}
        \centering
        \includegraphics[width=\textwidth]{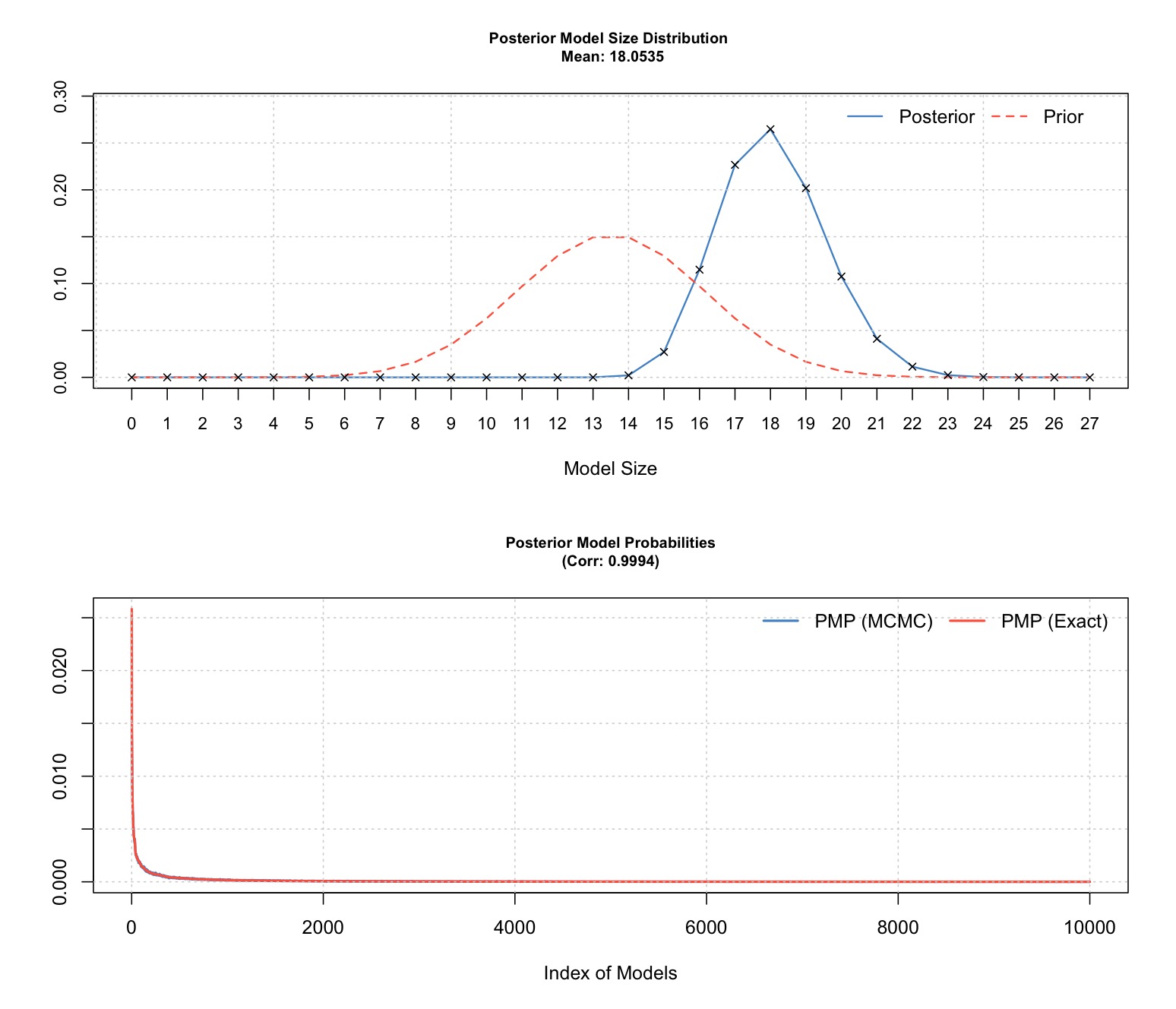}
        \caption{EBL and uniform priors}
        \label{fig:PMP_EBL_uniform}
    \end{minipage}

    % Add some vertical spacing between rows
    \vspace{0.3cm}

    % Second row of images
    \begin{minipage}[t]{0.48\textwidth}
        \centering
        \includegraphics[width=\textwidth]{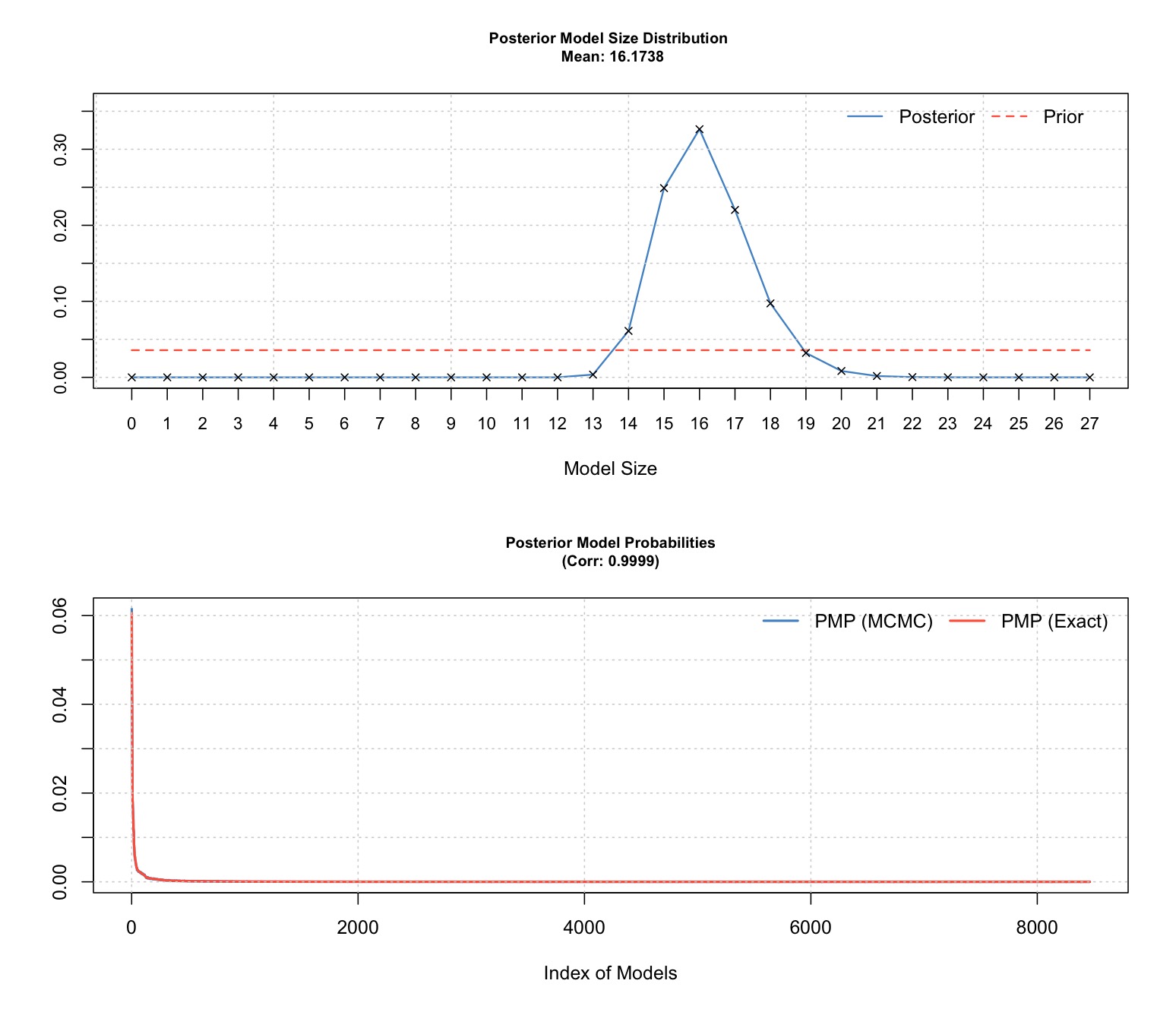}
        \caption{BRIC and random priors}
        \label{fig:PMP_BRIC_random}
    \end{minipage}
    \hfill
    \begin{minipage}[t]{0.48\textwidth}
        \centering
        \includegraphics[width=\textwidth]{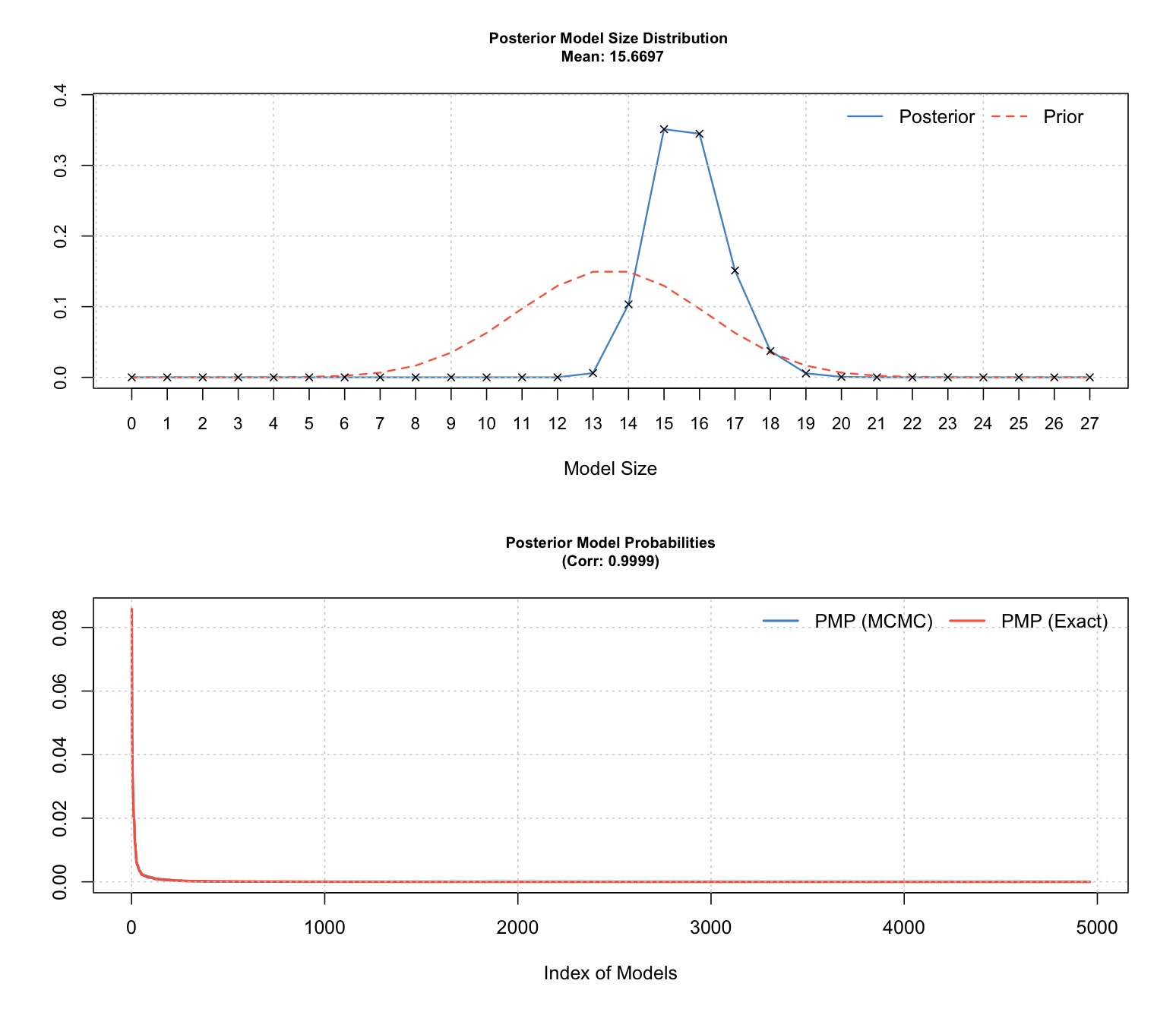}
        \caption{BRIC and uniform priors}
        \label{fig:PMP_BRIC_uniform}
    \end{minipage}
\end{figure}

\end{document}